\documentclass[sigchi]{acmart}

\usepackage{color}
\usepackage[normalem]{ulem}
\usepackage{enumitem}

\usepackage{xcolor}

\definecolor{revise}{RGB}{173, 216, 230}
\AtBeginDocument{%
  }

\newcommand{\toolName}[1]{\textit{HyperMOOC}}

\newcommand{\HL}[1]{\textcolor{black}{#1}}

\newcommand{\wl}[1]{\textcolor{black}{#1}}

\newcommand{\HLF}[1]{\textcolor{black}{#1}}

\newcommand{\sideicon}[1]{%
\hspace{4mm}
  \makebox[0pt][r]{\raisebox{-2mm}{\includegraphics[width=0.04\textwidth]{#1}}\hspace{2mm}}%
}

\begin{document}


\author{Li Ye}
\authornote{These authors contributed equally to this work.}
\affiliation{%
  \institution{Hangzhou Dianzi University}
  \city{Hangzhou}
  \country{China}}
\affiliation{%
\institution{Zhejiang University}
\city{Hangzhou}
\country{China}}
\email{li-ye@zju.edu.cn}

\author{Lei Wang}
\authornotemark[1]
\affiliation{%
  \institution{Hangzhou Dianzi University}
  \city{Hangzhou}
  \country{China}}
\email{leiwang@hdu.edu.cn}

\author{Lihong Cai}
\affiliation{%
  \institution{Hangzhou Dianzi University}
  \city{Hangzhou}
  \country{China}}
\email{lihongcai@hdu.edu.cn}

\author{RuiQi Yu}
\affiliation{%
  \institution{Hangzhou Dianzi University}
  \city{Hangzhou}
  \country{China}}
\email{ruiqiyu@hdu.edu.cn}

\author{Yong Wang}
\authornote{Corresponding author: Yong Wang and Zhiguang Zhou.}
\affiliation{%
  \institution{Nanyang Technological University}
  \city{Singapore}
  \country{Singapore}}
\email{yong-wang@ntu.edu.sg}

\author{Yigang Wang}
\affiliation{%
  \institution{Hangzhou Dianzi University}
  \city{Hangzhou}
  \country{China}}
\email{yigang.wang@hdu.edu.cn}

\author{Wei Chen}
\affiliation{%
\institution{Zhejiang University}
\city{Hangzhou}
\country{China}}
\email{chenvis@zju.edu.cn}

\author{Zhiguang Zhou}
\authornotemark[2]
\affiliation{%
  \institution{Hangzhou Dianzi University}
  \city{Hangzhou}
  \country{China}}
\email{zhgzhou@hdu.edu.cn}


\title{\toolName{}: Augmenting MOOC Videos with Concept-based Embedded Visualizations}

\begin{abstract}

Massive Open Online Courses (MOOCs) have become increasingly popular worldwide. However, learners primarily rely on watching videos, easily losing knowledge context and reducing learning effectiveness. We propose \toolName{}, a novel approach augmenting MOOC videos with concept-based embedded visualizations to help learners maintain knowledge context. Informed by expert interviews and literature review, \toolName{} employs multi-glyph designs for different knowledge types and multi-stage interactions for deeper understanding. Using a timeline-based radial visualization, learners can grasp cognitive paths of concepts and navigate courses through hyperlink-based interactions. We evaluated \toolName{} through a user study with 36 MOOC learners and interviews with two instructors. Results demonstrate that \toolName{} enhances learners' learning effect and efficiency on MOOCs, with participants showing higher satisfaction and improved course understanding compared to traditional video-based learning approaches.
\end{abstract}


\maketitle

\begin{figure*}
\centering
\begin{minipage}{1\textwidth}
    \centerline{\includegraphics[width=1\textwidth]{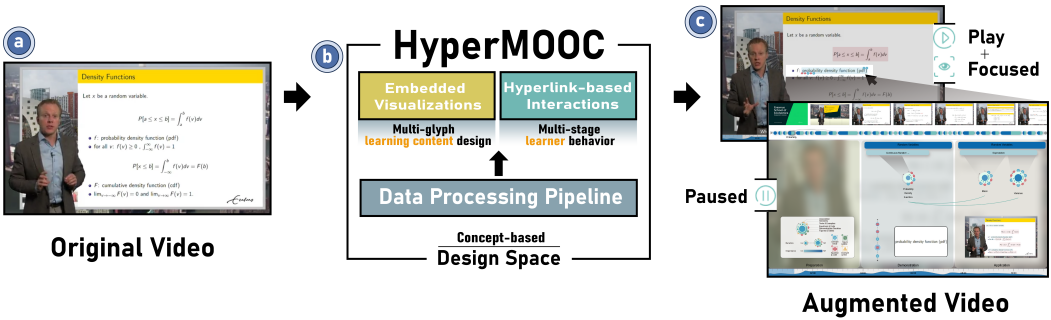}}
\end{minipage}
\vspace{-2mm}
\caption{
  Left: An original MOOC video. Middle: \toolName{} utilizes deep learning models to extract data from the original video, allowing learners to deepen their course understanding through direct interactions with elements in the video, facilitated by a concept-based design space, multi-glyph embedded visualizations, and hyperlink-based interactions. Right: The multi-stage augmented video highlights course organization and concept relationships.
}
\vspace{-4mm}
\label{fig:teaser}
\end{figure*}





\graphicspath{{icons/}{./}{figs/}{figures/}{pictures/}{images/}{./}} 





\section{Introduction}
With the rise of distance learning and online education platforms, Massive Open Online Courses (MOOCs) have continued to attract new learners over the past decade. 
However, learners often find it confusing to watch MOOC videos without the aid of additional knowledge\cite{milligan2016understanding} due to their lack of sufficient knowledge to understand the complex arrangement of the course and the diverse logical relationships of concepts \textcolor{black}{ - the fundamental ideas, theories, and principles that form the core of the course content.}
Existing methods of providing additional information, such as navigation menus~\cite{zhang2019scaffomapping,zhou2024conceptthread} and course overviews~\cite{rahman2023enhancing,schwab2016booc}, 
\HL{require learners to shift attention between different interfaces when explanatory materials or conceptual flows are placed outside the main video view. This increases cognitive load and disrupts conceptual continuity\cite{kizilcec2015instructor}.}
Thus, finding a way to seamlessly access additional data while watching a MOOC video is crucial for \textcolor{black}{effectively learning and understanding a} course.

\wl{Embedded visualizations present a promising opportunity to access additional data by directly displaying them within actual scenes, enabling seamless transitions between real-time viewing and on-demand exploration without diverting user's attention from MOOC videos.}
Various commercial products~\cite{CourtVision,VizLibero} and research systems~\cite{chen2023iball,chen2021augmenting} have leveraged embedded visualizations to augment videos, but these systems either do not focus on the educational domain or only utilize basic non-interactive elements.
Meanwhile, hypervideo~\cite{chambel2006hypervideo}, as an emerging interactive video technology, allows users to access additional information on-demand by embedding hyperlinks within the video. \HL{Existing survey feedback indicates that hypervideo is regarded as an effective means in MOOCs for promoting knowledge acquisition and enhancing learner engagement~\cite{milligan2016understanding}.}
However, to the best of our knowledge, there is still no work focused on hypervideo-based augmentation for MOOC videos.
\HL{MOOC videos are characterized by their short, modular format (often 5-15 minutes), high density of concepts, and the need to support diverse learners who learn asynchronously~\cite{pilli2016taxonomy}. These distinct characteristics create a gap, as generic hypervideo solutions lack the specialized capabilities for automated, curriculum-aware augmentation that MOOC contexts require.}

\HL{Understanding how to augment MOOC videos effectively requires examining both the content being taught and the learners engaging with it. Previous research has established that MOOC learning is fundamentally an interaction between two key components: the learning content \textit{(Object)} and the learner \textit{(Subject)}~\cite{blum2021stepping,lei2015advancing}. This object-subject framework has been widely adopted in learning content analysis~\cite{zawacki2018research,shih2005content}. However, augmenting MOOC videos within this framework presents unique challenges across both dimensions.
At the \textbf{learning content level}, existing systems~\cite{shih2005content} categorize video content into multiple hierarchical levels, but provide limited guidance on how to effectively augment these different types of data. This raises two critical questions: \textit{What kinds of data can be used to augment MOOC videos (Q1)}? \textit{What visual effects are used to augment these data (Q2)}?
At the \textbf{learner level}, MOOC learning is primarily learner-driven~\cite{blum2021stepping,lei2015advancing}, where individual learners exhibit diverse viewing behaviors and cognitive states. Research on cognitive load theory has demonstrated that excessive cognitive load can significantly impair learning outcomes~\cite{sweller1998cognitive,chen2017using}. This introduces two additional challenges: \textit{How to dynamically augment data based on learners' viewing behavior (Q3)}? \textit{How to augment data without increasing the cognitive load on learners while promoting their understanding (Q4)}?}

In this work, we aim to fill this gap by developing interactive embedded visualizations to help learners watch MOOC videos. 
Through systematic analysis of previous literature, learners' feedback, and expert interviews, we derived a concept-based design space~\cite{latham2020concept} to address the four aforementioned questions. 
The design space describes how to effectively augment MOOC videos in terms of both learning content and learner levels. The four dimensions of the design space (i.e., data type, visual effects, behavior level, and cognitive level) provide guidance for the augmentation of MOOC videos, 
offering new perspectives for facilitating learning in the online education domain.

Guided by the design space, we designed and implemented \toolName{}, a MOOC video augmentation tool that enhances learners' engagement and understanding of MOOC videos Fig.~\ref{fig:teaser}-c).
\toolName{} uses raw MOOC videos as input (Fig.~\ref{fig:teaser}-a), extracts the elements at each element level (Sec.~\ref{secIII.C}) through deep learning (DL) models, and implements a hyperlink-based interaction method to embed visualizations into MOOC video content (Fig.~\ref{fig:teaser}-b).
Our interaction flow supports three stages—``\textit{Play}'', ``\textit{Focused}'', and ``\textit{Paused}''—corresponding to continuous playback, focused attention on specific content, and pausing for deeper engagement. Using keyboard-mouse interaction, learners can seamlessly transition between these stages, balancing explanatory and exploratory aspects within MOOC video learning.
Guided by the multi-stage methodology of knowledge acquisition~\cite{AboutLearning}, we implemented concept-based glyphs to visually encode various types of basic data elements (figures \& tables, equations \& codes, examples \& tests).
A user study with 36 learners demonstrated the overall usability of \toolName{}. Comparing three conditions (RAW, AUG, FULL), participants agreed that the FULL version can more effectively promote course understanding. We further collected feedback from two educational psychologists and discussed limitations and future research directions.
The corpus, created videos, and other materials can be found in \url{https://hypermooc.github.io/HyperMOOC/}.
In summary, \textcolor{black}{the contributions of this paper can be summarized as follows:}
\vspace{-1mm}
\begin{itemize}
	\setlength{\itemsep}{0pt}
	\setlength{\parsep}{0pt}
	\setlength{\parskip}{0pt}
  \item \textbf{Design Space.} We derive a two-level, four-dimensional concept-based design space for augmenting MOOC videos from existing literature on video augmentation and online learning, combined with analyses from learners and experts.
  
  \item \textbf{Fast Prototyping Tool.} We design and implement \toolName{}, a fast prototyping tool for augmented MOOC videos that combines multi-glyph designs for concept demonstration and hyperlink-based interaction methods based on learners' MOOC video learning behaviors.

  \item \textbf{Evaluation.} A user study conducted with \textcolor{black}{36} learners and expert interviews show that \toolName{} can effectively facilitate online learners in conveniently and quickly learning from MOOC videos.
  
\end{itemize}


\section{Related Work}

\textbf{Learning Content Visual Analysis.}
Previous research has categorized learning content visual analysis systems into three levels: object-level, event-level, and conclusion-level\cite{shih2005content}, aiming to aid students in acquiring knowledge more effectively in MOOC learning. 


\textit{\uline{Object-level}} analysis focuses on indexing, retrieving, and browsing key concepts of online courses. Representative works include extracting index words from audio transcriptions\cite{haubold2004analysis}, applying OCR and keyword recognition to generate searchable term lists from lecture slides\cite{adcock2010talkminer}, and unsupervised keyphrase extraction methods\cite{albahr2019semkeyphrase}, all reducing the difficulty of obtaining key concepts.


\textit{\uline{Event-level}} analysis focuses on identifying patterns between concepts and analyzing their interrelationship during concept transitions. 
\HL{For instance, ConceptScape\cite{liu2018conceptscape} identifies transitions such as \textit{algorithm} $\rightarrow$ \textit{time complexity} $\rightarrow$ \textit{optimization}, capturing not only temporal order but also causal and hierarchical relationships between concepts.}
Concept mapping plays a crucial role in knowledge representation and assessment\cite{chang2001learning,oliver2009investigation}. To reduce learners' effort in crafting concept maps, tools like CmapTools\cite{canas2014concept} and ScaffoMapping\cite{zhang2019scaffomapping} provide auxiliary functions to enhance mapping quality.


\textit{\uline{Conclusion-level}} analysis focuses on providing course summaries to help learners grasp core concepts. Works in this area segment lecture videos with visual hints\cite{haubold2005augmented}, provide content overviews through keyframes\cite{zhao2016new}, and propose automated segmentation methods to improve navigation efficiency\cite{rahman2023enhancing}.

\HL{Although the three levels of analysis offer complementary perspectives on learning content, existing systems typically treat them in isolation, lacking an integrated framework that connects object-, event-, and conclusion-level insights in a coherent design space, which limit learners’ ability to explore complex course materials holistically.}

\noindent
\textbf{Online Learning Video Visualizations.}
Numerous studies have explored diverse methods to visualize online learning videos, aiming to enhance educators' and learners' understanding and engagement with educational content, \HL{leading to understanding the value of online learning data from a visual analytics perspective\cite{emmons2017mooc}}.


Visual analysis of clickstream data is crucial for tracking student progress. Systems such as PeakVizor\cite{chen2015peakvizor}, VUSphere\cite{he2018vusphere}, and ViSeq\cite{chen2018viseq} employ various visualization techniques—clock-like glyphs, spherical layouts, and bidirectional chord diagrams—to analyze learner behaviors and correlate learning sequences with performance.


In terms of visualizing learning content, several works present concepts to make learning more engaging. Zhao et al.\cite{zhao2017novel} presented keyframe summaries in a comic book style. \HL{Torre et al.\cite{torre2022video} proposed augmented transcripts and concept maps that dynamically highlight relevant segments, facilitating cognitive focus during video-based learning.} Zhou et al.\cite{zhou2024conceptthread} proposed ConceptThread, presenting course concepts in a thread view to facilitate understanding of concept relationships.

\HL{Existing work has yet to explore how visualizations can be embedded directly into MOOC videos to maintain learners’ contextual awareness without diverting attention away from the primary learning surface.  To address this gap, our study aims to investigate how integrated video-embedded visualizations can support continuous situational awareness learning during video viewing.}


\noindent\textcolor{black}{
\textbf{Hypervideo,}
as an extension of hypertext to video media, has gained significant attention in educational contexts. Based on the characteristics summarized by Sauli et al.\cite{sauli2018hypervideo}, we focus on three key features of hypervideo: dynamism, control features, and hyperlink.}

\textcolor{black}{\textit{\uline{The dynamic nature}} of hypervideo transforms static information into engaging and motion-based content. For instance, Chambel et al.\cite{chambel2006hypervideo} demonstrated that hypervideo could visualize static dimensions as dynamic ones, thus promoting comprehension. In MOOCs, this dynamism is particularly beneficial, allowing viewers to access additional information about video clips through hyperlinks.}

\textcolor{black}{\textit{\uline{Control features}} in hypervideo systems empower users to navigate and interact with video content more flexibly. HyperMeeting\cite{girgensohn2015hypermeeting} exemplifies this by offering extensive control over video playback, allowing participants to initiate playback for all current participants, navigate between previously meetings, and pause or skip recorded video streams.}

\textcolor{black}{\textit{\uline{Hyperlink}} is central to hypervideo, enabling the creation of interconnected video content and additional information. As seen in TrACE\cite{dorn2015piloting}, which allows viewers to annotate and embed discussions at specific locations and times in videos. Tiellet et al.\cite{tiellet2010design} created MultiLinks, which allows viewing surgeries organized in steps, linked to corresponding times in the hypervideo, providing effective learning support in veterinary surgery education.}

\HL{Although these systems demonstrate how navigation, annotation, and dynamic visualization can enrich educational video interactions, they typically treat these functions in isolation and rarely leverage them to uncover learners’ cognitive pathways or support conceptual understanding in MOOCs.
{\toolName{} advances this paradigm by integrating conceptual visualizations, multi-symbol design, staged interactions, cognitive-path displays, and hyperlink-based navigation. This integrated design deepens learners’ understanding of MOOC content and knowledge structures while enhancing overall learning efficiency and experience.}}

\noindent\textcolor{black}{
\textbf{Video Interaction Methods.}
Video interaction techniques in augmented visualization aim to enhance user experience and analytical efficiency. Common methods include keyboard-mouse interaction, eye-tracking, and Human-in-the-Loop (HITL) approaches. Keyboard-mouse interactions, such as elastic timing sliders \cite{hurst2004advanced} and hierarchical navigation \cite{girgensohn2011adaptive}, offer intuitive control over video playback and navigation. Eye-tracking has emerged as a promising approach for personalized interactions, enabling content-aware video editing \cite{moorthy2020gazed} and optimized video repositioning \cite{rachavarapu2018watch}. HITL mechanisms play a crucial role in improving user experience and interaction efficiency \cite{sacha2016human}, as demonstrated by systems like VideoPro \cite{he2023videopro} and VisCommentator \cite{chen2021augmenting}, which allow for dynamic content selection and iterative refinement of visual effects. Our system currently employs keyboard-mouse and HITL interactions, with potential future integration of eye-tracking. We propose a hyperlink-based interaction approach to enhance learners' understanding of content, reduce cognitive load, and improve learning efficiency and outcomes.}

\section{Design Space for Augmenting MOOC Videos}

To understand how to augment MOOC videos, we derive design practices from a literature review, user interviews, and expert feedback to characterize augmentations of the object (learning content) and subject (learner) of MOOC learning.

\vspace{-6mm}
\HL{
\subsection{Methodology}
We employed a triangulated methodological approach combining literature review, user interviews, and expert consultations to establish a rigorous foundation for our design space, ensuring that our design choices are grounded in theoretical understanding, empirical evidence from learners, and practical insights from domain experts.}

\begin{figure*}[t]
    \centering
    \includegraphics[width=\linewidth]{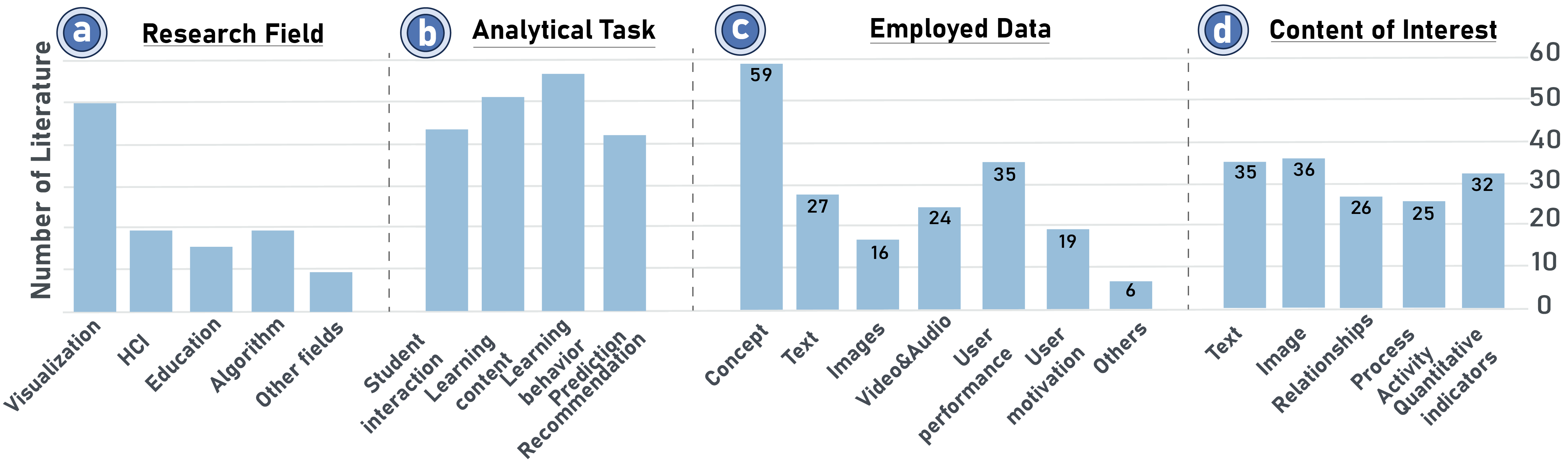}
    \vspace{-6mm}
    \caption{\HLF{Literature review result, including the number of literature on a) research filed, b) analytical task, c) employed data, and d) content of interest.}}
    \vspace{-6mm}
    \label{Literature}
\end{figure*}

\noindent
\textbf{\HL{Literature Review.}}
\HL{We conducted a systematic literature review using both search-driven and reference-driven methods. The search employed keywords ``online learning'', ``education videos'', ``MOOC videos'', ``learning content analysis'', and ``e-learning analysis'' across IEEE Xplore, ACM Digital Library, and Google Scholar, covering publications from 2010 to 2025. We focused on high-impact venues including conferences (IEEE VIS, ACM CHI, ACM L@S, ACM LAK) and journals (IEEE TVCG, Computer Graphics \& Applications, Computer \& Education). Following initial searches, we recursively examined references within identified papers. Papers were included if they focused on MOOC/online educational videos, presented analytical or visualization techniques, or investigated learner behaviors in video-based online education. This process yielded 74 representative papers (Fig.~\ref{Literature}-a).}

\HL{Three co-authors collaboratively coded these papers along four dimensions: research field, analytical task, employed data, and content of interest. Each author independently reviewed 1/3 of the papers while cross-validating another 1/3 reviewed by others, following categorization criteria from prior work~\cite{kui2022survey,zhang2022towards}. Papers could receive multiple labels when applicable, and discrepancies were resolved through iterative discussions until consensus was reached.}



\noindent
\textbf{\HL{User Interview.}}
\HL{Given that MOOC learning is primarily learner-driven~\cite{lei2015advancing,blum2021stepping}, we conducted semi-structured interviews with 10 university students (4 computer science, 3 finance, 3 art design majors) who had used MOOCs actively for over 3 years and completed at least 5 courses on major platforms. Participants were recruited through university mailing lists. The 45-60 minute face-to-face interviews investigated: (1) how participants approach and process content when watching MOOC videos, including viewing habits and comprehension tactics, and (2) specific points where they encounter difficulties or require additional support. The interview protocol was developed based on findings from our literature review to ensure alignment with documented challenges.}

\HL{During interviews, participants browsed actual MOOC platforms to demonstrate typical learning behaviors and articulate their thought processes in real-time. One co-author documented each session, recording both verbal responses and observed behaviors. Interview data were analyzed using iterative coding to identify recurring patterns and learner needs that informed our design requirements.}



\noindent
\textbf{\HL{Expert Consultation.}}
\HL{We conducted structured consultation sessions with two domain experts who bring complementary perspectives on MOOC pedagogy and analytics. E1 is a MOOC instructor with 3 years of experience developing online lecture videos on Coursera and edX, providing practical insights from teaching thousands of students. E2 is an educational researcher with 10 years of experience in MOOC analytics, offering theoretical grounding and empirical evidence from large-scale learning analytics studies.}

\HL{The face-to-face consultation sessions (\~2 hours each) followed a systematic protocol: First, we presented our literature review findings to establish shared understanding of the research landscape. Second, we shared synthesized results from student interviews, focusing on learner behavioral patterns and identified pain points. Third, we solicited expert perspectives on common difficulties in online learning, evaluation of our proposed design dimensions, and refinements based on their domain knowledge. This expert consultation served as critical triangulation, validating that our design choices integrated theoretical foundations, empirical evidence from learners, and practical wisdom from experienced practitioners and researchers.}

\subsection{Concept-based Design Space}

\HL{Instructors integrate and organize concepts into videos, and learners complete learning by watching the entire video.
This is the most basic delivery mode in online learning~\cite{greenhow2022foundations}. The necessity of concept-based augmentation is grounded in established theories of learning and cognition. Ausubel's meaningful learning theory~\cite{ausubel1978educational} posits that ``the most important single factor influencing learning is what the learner already knows,'' emphasizing that learners construct understanding by relating new information to existing conceptual frameworks. Building on this foundation, Erickson's concept-based learning model~\cite{erickson2007concept} demonstrates that concepts, as abstract, timeless, and universal ideas, serve as the cognitive bridge enabling knowledge transfer across contexts, a critical requirement for effective MOOC learning~\cite{latham2020concept,lasater2009influence}. Unlike isolated facts that remain ``locked in time and place,'' concepts facilitate deeper understanding through term ``synergistic thinking'', the cognitive interplay between factual and conceptual levels of mental processing~\cite{erickson2007concept}.}

\HL{This theoretical foundation is empirically supported by our literature review, which revealed that 79.7\% (59 out of 74) of papers analyzing MOOC videos focus on concepts as the fundamental unit of analysis (Fig.~\ref{Literature}-b,c), employing different representations of concepts (Fig.~\ref{Literature}-d) to accomplish diverse analytical tasks. Further discussions with domain experts confirmed that concept-based analysis consistently leads to enhanced learner understanding and course progress. The pedagogical benefits are substantial: concept-based approaches promote transferable understanding rather than rote memorization, enable learners to see connections between disparate topics, and support the construction of coherent mental models, all essential for self-directed MOOC learning where instructional scaffolding is limited~\cite{erickson2007concept,latham2020concept,lasater2009influence}.}

\HL{Thus, embedding visualizations in MOOC videos must be concept-based, enabling learners to engage with the MOOC video thoroughly from the bottom up. This approach aligns well with MOOC video augmentation's unique characteristics and requirements.}

\subsection{Learning Content-level Design}
\label{secIII.C}
We start augmenting MOOC videos from the \textit{Object} of MOOC learning (learning content), and we focus on two main questions, i.e., \textit{what data in MOOC videos can be augmented (Q1)?} and \textit{how can we visually augment these data (Q2)?}
 \textcolor{black}{Based on these questions and inspired by Chen et al.'s design space for augmented videos~\cite{chen2021augmenting}, we characterize these two questions using Data Type and Visual Effect.
}

\noindent\textbf{Dimension I: Data Type.}
\HL{The derivation of this dimension triangulates three sources: literature on video augmentation establishing the need to balance explanatory and exploratory approaches~\cite{chen2023iball}, our observation that 79.7\% of analyzed papers distinguish between content-level and relationship-level data, and user interviews where 8 out of 10 participants explicitly differentiated ``understanding what's on the slide'' versus ``understanding how concepts connect.'' Given that MOOC learning is multi-stage~\cite{AboutLearning}, learners first acquire basic concepts before applying knowledge in deeper exploration. We thus identified a semantic hierarchy: explanatory content exhibits low semantic complexity (text, figures, equations directly embedded), while exploratory data carry high semantic complexity (concept relationships, importance indicators requiring deeper cognitive processing). Expert E1 confirmed: ``students first need to see what's there before understanding how it all fits together.'' This semantic stratification provides the foundation for categorizing MOOC video data from low to high levels:}

\sideicon{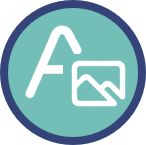}%
\textcolor{black}{\textit{\uline{Element Level}} includes the most general representation of the concept, which can be obtained directly from the original video. 
Learners generally use this level of content to gain the most general knowledge of the course. 
Based on previous literature~\cite{yadav2016vizig,chen2015effects,chorianopoulos2013usability} and our review (Fig.~\ref{Literature}-d), data at this level generally consists of basic elements describing the concept and auxiliary elements provided by the instructor. Specifically, these include: (1) Basic elements—such as text, figures, tables, equations, and code blocks; and (2) Auxiliary elements—such as teacher images, subtitles, tests, and examples. For instance, in a statistics course on MOOC, a textual definition of "mean" displayed alongside an illustrative table constitutes Element Level data. Augmenting a video solely at this level is primarily \textbf{for explanatory purposes}.}

\sideicon{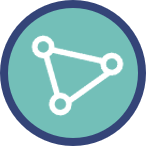}%
\textit{\uline{Event Level}} involves higher-level abstractions of concepts,
which are generally difficult for learners to access directly from videos.
\HL{The extraction of key concepts and the establishment of their structured relationships to enhance learners' understanding has also garnered significant attention from learners\cite{zhang2019scaffomapping,zhao2017novel}. Donald\cite{donald1983knowledge} categorized the relationships between concepts into three types: associative, functional, and structural, which are further mapped to Sequence, Association, Similarity, and Inclusion to better express the relationships between concepts in MOOC videos.
 Providing additional presentation of this data is particularly effective for logic-driven and concept-intensive courses~\cite{zhou2024conceptthread}. Thus, we categorize this data into concept attributes and relationships.} Attributes include \textit{duration, importance, and delivery style (i.e., whiteboard annotations, slide-based presentations, direct lecture delivery)} and relationships consist of \textit{association, inclusion, and similarity}. Augmenting data at this level makes videos \textbf{both explanatory and exploratory}.

\sideicon{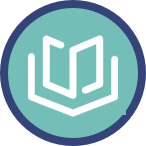}%
\textcolor{black}{\textit{\uline{Conclusion Level}} presents course-level data, enhancing learners' overall grasp of the course. This type of data explains the overall composition of the course and outlines the learning steps, such as the \textit{course's type, structure, and important time nodes}.
Most current research based on learning and content analysis focuses on this level, using navigation menus~\cite{zhou2024conceptthread,zhao2017novel} and course overview~\cite{rahman2023enhancing,schwab2016booc} to facilitate quick learning.
Therefore, augmenting data (e.g., temporal structure, concept hierarchy) at this level is primarily \textbf{for exploratory purposes}, providing learners with additional information.}

 \textcolor{black}{\toolName{} leverages data across all semantic levels, from element-level basic concepts to conclusion-level course structures. This multi-level approach ensures that learners can access both explanatory and exploratory content, facilitating a comprehensive understanding of the course material.}

\noindent\textbf{Dimension II: Visual Effect.}
\HL{Having established what to augment, this dimension addresses how. In the data video domain, data may be presented as graphical marks~\cite{amini2017authoring}, combined into information-rich frames to convey insights. User interviews validated this framework as participants naturally distinguished ``seeing additional information'' versus ``being guided to look at something,'' while E2 characterized it as ``showing versus highlighting.'' Drawing from visualization grammar~\cite{amini2017authoring} and video augmentation literature~\cite{Chen2023Sporthesia,chen2021augmenting,chen2023iball}, we categorize visual effects as: (1) \textit{graphical marks} providing persistent information overlays, and (2) \textit{video effects} dynamically redirecting attention.}

\noindent
\begin{itemize}[leftmargin=*]
    \setlength{\leftmargin}{0pt}
	\setlength{\itemsep}{0pt}
	\setlength{\parsep}{0pt}
	\setlength{\parskip}{0pt}
 
\item \textit{\uline{Embedded Representation}} are static visual elements directly integrated into the original video.
Common augmented video work~\cite{chen2023iball} usually uses marks such as text and wireframes to explain the content, while more sophisticated data videos~\cite{amini2017authoring} incorporate complex visualizations like explanatory figures and basic charts. Therefore, we categorize basic graphical marks into
\hspace{-1mm}
\raisebox{-1.3mm}{\includegraphics[scale=0.55]{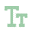}}
\hspace{-1mm}
\textit{Text}, 
\hspace{-1mm}
\raisebox{-1.3mm}{\includegraphics[scale=0.55]{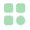}}
\hspace{-1mm}
\textit{Figures}, and 
\hspace{-1mm}
\raisebox{-1.3mm}{\includegraphics[scale=0.55]{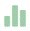}}
\hspace{-1mm}
\textit{Charts}. 
Further, based on extensive expert feedback and E1's experience in MOOC teaching, 
\hspace{-1mm}
\raisebox{-1mm}{\includegraphics[scale=0.6]{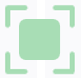}}
\hspace{-1mm}
\textit{Annotations} are also considered a critical graphical mark. Instructors use drawing-board to explain video content in real-time, which is deemed as an important teaching method~\cite{chorianopoulos2013usability}.

\vspace{1mm}
\item \textit{\uline{Video Effects}} are dynamic visual elements added to videos.
Unlike graphical marks, video effects tend to affect the viewer's emotion and attention.
 \textcolor{black}{In some ways, these effects resemble the attention cues discussed in the study by Amini et al~\cite{amini2017authoring}. }
Typical post-production of videos is done by adding
\hspace{-1mm}
\raisebox{-1.3mm}{\includegraphics[scale=0.52]{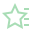}}
\hspace{-1mm}
\textit{Animation} and
\hspace{-1mm}
\raisebox{-1.0mm}{\includegraphics[scale=0.52]{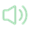}}
\hspace{-1mm}
\textit{Sound Effects}
to increase the experience during playback.
However, E2 points out that visual effects for MOOC videos should be designed to provide an in-depth, multi-dimensional learning experience.
Inspired by this and suggestions from MOOC video production~\cite{sauli2018hypervideo}, we provide additional video effects:
\hspace{-1mm}
\raisebox{-1.3mm}{\includegraphics[scale=0.52]{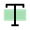}}
\hspace{-1mm}
\textit{Highlight} and
\hspace{-1mm}
\raisebox{-1mm}{\includegraphics[scale=0.52]{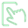}}
\hspace{-1mm}
\textit{Interactive Elements}.
Highlight effects emphasize key information, while Interactive Elements allow viewers to engage directly with the video content.

\end{itemize}

 \textcolor{black}{Our system implements a range of visual effects, including embedded representations (e.g., multi-glyph designs for different knowledge types) and video effects (e.g., highlights and interactive elements). These visual augmentations are carefully designed to enhance the learner's engagement with the content without overwhelming them.}

\vspace{-3mm}
\subsection{Learner-level Design}
MOOC video learning is mainly learner-driven~\cite{lei2015advancing,blum2021stepping}. Therefore, we focus on the behavioral patterns of the \textit{Subject} of MOOC learning (learner), i.e., \textit{what kind of behaviors learners generally engage in for acquiring knowledge (Q3)?} and \textit{what hinders/facilitates their understanding during the process of acquiring knowledge (Q4)?} We identify two design dimensions: Behavior Level and Cognitive Level.

\noindent\textbf{Dimension III: Behavior Level.}
\HL{This dimension emerged from a critical observation across our data: while instructors design MOOC videos with temporal linearity, learner behavior is non-linear and interactive. All 10 interview participants demonstrated varied engagement patterns—pausing, rewinding, and skipping—validated by literature on learner behavior~\cite{ozan2016video,shi2015vismooc,sinha2014your} documenting consistent interaction patterns beyond sequential viewing. Expert confirmation strengthened this finding: E1 noted ``students rarely just watch straight through—they pause, they jump around,'' while E2's analytics showed ``video interaction patterns strongly predict learning outcomes.'' Synthesizing literature, interview observations, and expert validation, we derived three fundamental behavioral states capturing how learners actively control their learning process from passive consumption to active exploration:}

\sideicon{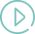}%
\textit{\uline{Play}} denotes the first step of learning, where the learner follows the video through the instructor's preset learning process. At this stage, learners tend to learn the most general, \textbf{explanatory} content.

\vspace{1mm}
\sideicon{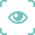}%
\textcolor{black}{\textit{\uline{Focused}} refers to the intermediate process of learning. Our user interviews showed that learners often moved their mouse over content being demonstrated by the instructor. Informed by E1, we observed that learners tend to spend more time focusing on difficult content, aligning with Miller's hypothesis of eye-movement cognitive engagement~\cite{miller2015using}. \wl{While eye-tracking provides more precise attention measures, we adopt a \textit{mouse-hover-based focus proxy} considering ecological validity: most MOOC learning occurs on personal devices without eye-trackers, and mouse cursor movements correlate with visual attention~\cite{huang2011no}.} This approach allows us to infer areas of focus and trigger relevant augmentations during the Focused stage.
Learners typically deepen their understanding of \textbf{explanatory} content and seek out \textbf{exploratory} information at this stage.}

\sideicon{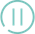}%
\textit{\uline{Paused}} stage tends to promote the final understanding of concepts.
Learners agreed that when the concepts they encountered were sufficiently difficult and complex, they often paused to explore them further.
According to a previous clickstream analysis~\cite{shi2015vismooc,sinha2014your}, ``seek'' and ``rate change'' tend to be learners' favorite actions.
Moreover, the learning process recorded by one co-author shows that beyond videos, some learners preferred to use search engines such as Google, Wikipedia, or ChatGPT to gain a broader understanding of concepts, thus obtaining \textbf{exploratory} content for concept cognition.

\sideicon{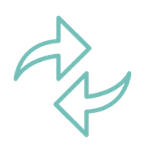}%
\textit{\uline{Seek}} occurs mainly in the \textit{Paused} stage, where learners drag the video timeline from one point to another to find supplementary content for concept explanation. 
However, learners often struggle to find the exact time point, and it usually takes several \textit{Seek} operations.

 \textcolor{black}{The \textbf{Behavior Level} interacts dynamically with the \textbf{Data Type} and \textbf{Visual Effect}. }
Learner behavior usually follows a staged cycle. Initially, they acquire broad insights by \textit{Play} the video, then engage in \textit{Focused} to track the instructor's teaching path or examine the content of interest, and ultimately \textit{Paused} for an in-depth understanding of the concepts.
During the \textit{Paused} stage, \textit{Seek} is the predominant action, aiding learners in revisiting content or facilitating rapid learning.

\noindent\textbf{Dimension IV: Cognitive Level.}
\HL{While prior dimensions emerged from literature and user data, this was explicitly proposed by both domain experts as a critical perspective. During consultations, E1 and E2 emphasized cognitive load management—E1 noting ``adding more information can sometimes make things worse,'' and E2 citing research linking cognitive load to MOOC completion rates. This directed us to cognitive load theory~\cite{sweller1998cognitive}. We categorize loads by their sources and instructional implications, distinguishing between unavoidable content difficulty, processing overhead from poor presentation, and productive cognitive effort for schema construction:}

\noindent
\begin{itemize}[leftmargin=*]
    \setlength{\leftmargin}{0pt}
	\setlength{\itemsep}{0pt}
	\setlength{\parsep}{0pt}
	\setlength{\parskip}{0pt}

\item \textit{\uline{Intrinsic Load}} denotes the cognitive load resulting from the difficulty of the course itself, which does not vary with the \textit{Object} and \textit{Subject} of learning. 
For example, solving a differential equation has a greater intrinsic load than calculating 2 + 2.
General cognitive augmentation methods do not optimize the cognitive load for this category.

\item \textit{\uline{Extraneous Load}} is generated by the manner in which information is presented to learners and is under the control of instructional designers~\cite{sweller1998cognitive}.
This burden can be attributed to the design of instructional materials, typically expressed as the organization of the learning \textit{Object} (learning content).
Therefore, learning materials should be designed to minimize extraneous load~\cite{ginns2006integrating}, suggesting that MOOC video augmentation should avoid overly complex visualizations.

\item \textit{\uline{Germane Load}} involves the mental model's processing, construction, and automation~\cite{sweller1998cognitive}, primarily attributed to the \textit{Subject} of learning (learner).
E2 suggests that reducing this load in MOOC video learning should depend on learners' access to instructional aids to facilitate understanding, such as high-semantic course information (e.g., concept overview or relationships).
User interviews also highlighted learners' preferences for seeking and searching during learning. 
Thus, additional information should be easily accessible.

\end{itemize}

 \textcolor{black}{By considering the different types of cognitive load, \toolName{} aims to minimize extraneous load through clear and intuitive visualizations and reduce the germane load by providing easily accessible additional information.}


\vspace{-2mm}
\section{HyperMOOC}
Based on the design space, we design \toolName{}, a fast prototyping tool for augmenting MOOC videos. \toolName{} is a web application implemented using Flask, VueJS, and D3.js.

\vspace{-1mm}
\subsection{Design Requirements}
We further derived the design requirements of \toolName{} based on the design space, previous research on augmented video and online learning~\cite{chen2021augmenting,zhou2024conceptthread,shi2015vismooc}, and weekly meetings with two domain experts (E1 \& E2, consistent with expert interviews) over the past 5 months. 
\HL{Notably, the experts' insights were grounded in systematic observations of learner behaviors from discussion forums and post-class feedback throughout their teaching experience. }
Besides, our design targets standard desktop environments because most current MOOC platforms are mainly accessed by online learners via computers.
Ultimately, we grouped them into three categories: for explanatory purposes (
\hspace{-1.7mm}
\raisebox{-1.6mm}{\includegraphics[scale=0.52]{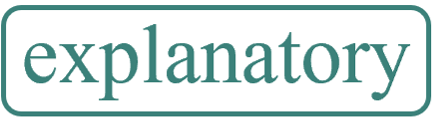}}
\hspace{-1.7mm}
), for exploratory purposes (
\hspace{-1.7mm}
\raisebox{-1.6mm}{\includegraphics[scale=0.52]{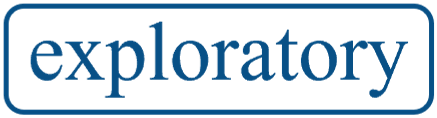}}
\hspace{-1.7mm}
) and for both explanatory and exploratory purposes (
\hspace{-1.7mm}
\raisebox{-1.5mm}{\includegraphics[scale=0.52]{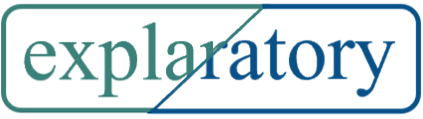}}
\hspace{-1.7mm}
).

\vspace{-1mm}
\begin{enumerate}[label=\textbf{R\arabic*},leftmargin=*]
\setlength{\itemsep}{0pt}
\setlength{\parsep}{0pt}
\setlength{\parskip}{0pt}


\item \raisebox{-1.5mm}{\includegraphics[scale=0.52]{./icon-explanatory.png}}\hspace{-0.5mm}
\textbf{Provide an overview of concepts and enable real-time tracking.}
Existing analyses of learning content~\cite{zhou2024conceptthread,schwab2016booc} provide concept overviews as the initial learning step, and augmented videos commonly employ \textit{Highlight} effects for real-time augmentation~\cite{chen2021augmenting,chen2023iball}. 
\HL{E1 observed that learners frequently posted forum questions like ``\textit{where are we in the course?}'' and post-class feedback confirmed that nearly 70\% of learners reported difficulty tracking course structure. Thus, enabling learners to understand general course content and track their current concept is the most fundamental MOOC video augmentation.}

\item \raisebox{-1.5mm}{\includegraphics[scale=0.52]{./icon-explanatory.png}}\hspace{-0.5mm}
\textbf{Display elements that describe the concept.}
According to our design space, concepts have a variety of representations.
\HL{E1 noted learners' common forum complaints about difficulty locating ``\textit{which slide had 
that example,}'' and mentioned frequently ``\textit{need pausing to find specific content.}'' 
Explicitly displaying this information can help learners quickly access concept-related material and the concept demonstration flow, which is crucial for promoting self-perceived learning.}

\item \raisebox{-1.5mm}{\includegraphics[scale=0.52]{./icon-exploratory.png}}\hspace{-0.5mm}
\textbf{Show the relationship between concepts and enable rapid navigation.}
Experts agree that the relationships between concepts are the most critical elements to understanding the course. ``\textit{Learners' exploration in courses is often guided by relationships, it's important to demonstrate these relationships and provide interactive navigation,}'' E2 commented. E1 echoed this, highlighting the significance of interactive video navigation.


\item \raisebox{-1.5mm}{\includegraphics[scale=0.52]{./icon-both.png}}\hspace{-0.5mm}
\textbf{Offer support for multi-stage learning behaviors.}
\HL{Given multiple learning behaviors observed by both experts, the augmented design should support different stage-related activities through seamlessly embedded visualizations.
Specifically, the system should provide real-time, low-interruption augmentation during video playback (Play and Focused stages) and enable on-demand exploration when learners pause, all within the video canvas to maintain spatial continuity and minimize attention shifts.}


\item \raisebox{-1.5mm}{\includegraphics[scale=0.52]{./icon-both.png}}\hspace{-0.5mm}
\textbf{Ensure augmented content does not obstruct the learner's learning.}
After several rounds of discussions with experts on the prototype tool, they agreed that the purpose of augmentation should be to facilitate learning rather than cause confusion. 
E1 believed that ``\textit{augmented content should primarily be video-based, rather than multi-view,}'' and E2 added, ``\textit{visual designs used for augmentation should not be overly complex.}''

\end{enumerate}

\begin{figure}[h]
    \centering
    \includegraphics[width=\linewidth]{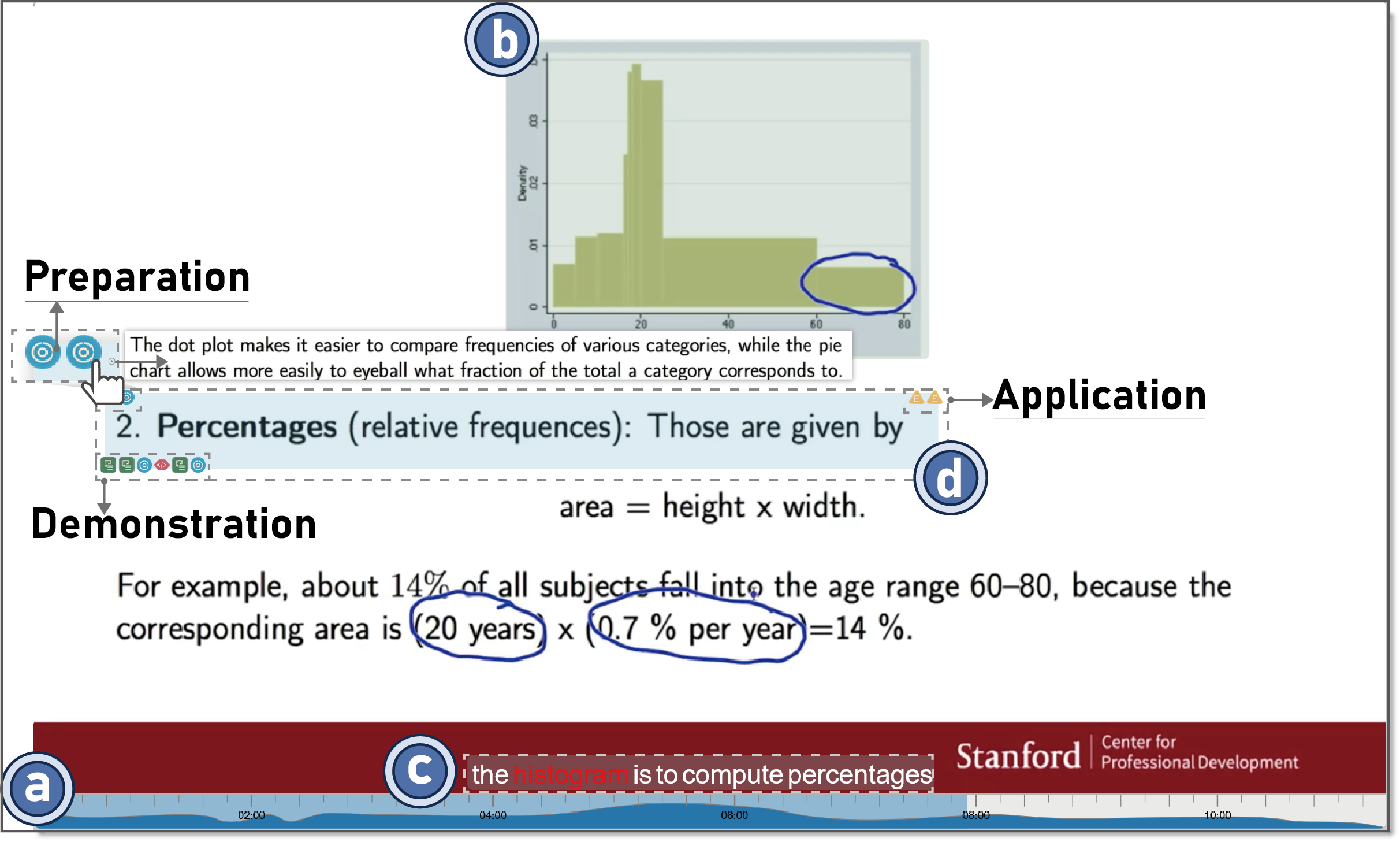}
 
    \caption{\HL{The design of \textit{Play} (a,b,c) and \textit{Focused} (d) stage of \toolName{}, Alex is currently viewing the original content of concept ``Percentages'' in ``Preparation'' stage.}}
    \vspace{-4mm}
    \label{play-focused}
\end{figure}

\begin{figure*}[t]
    \centering
    \includegraphics[width=0.95\linewidth]{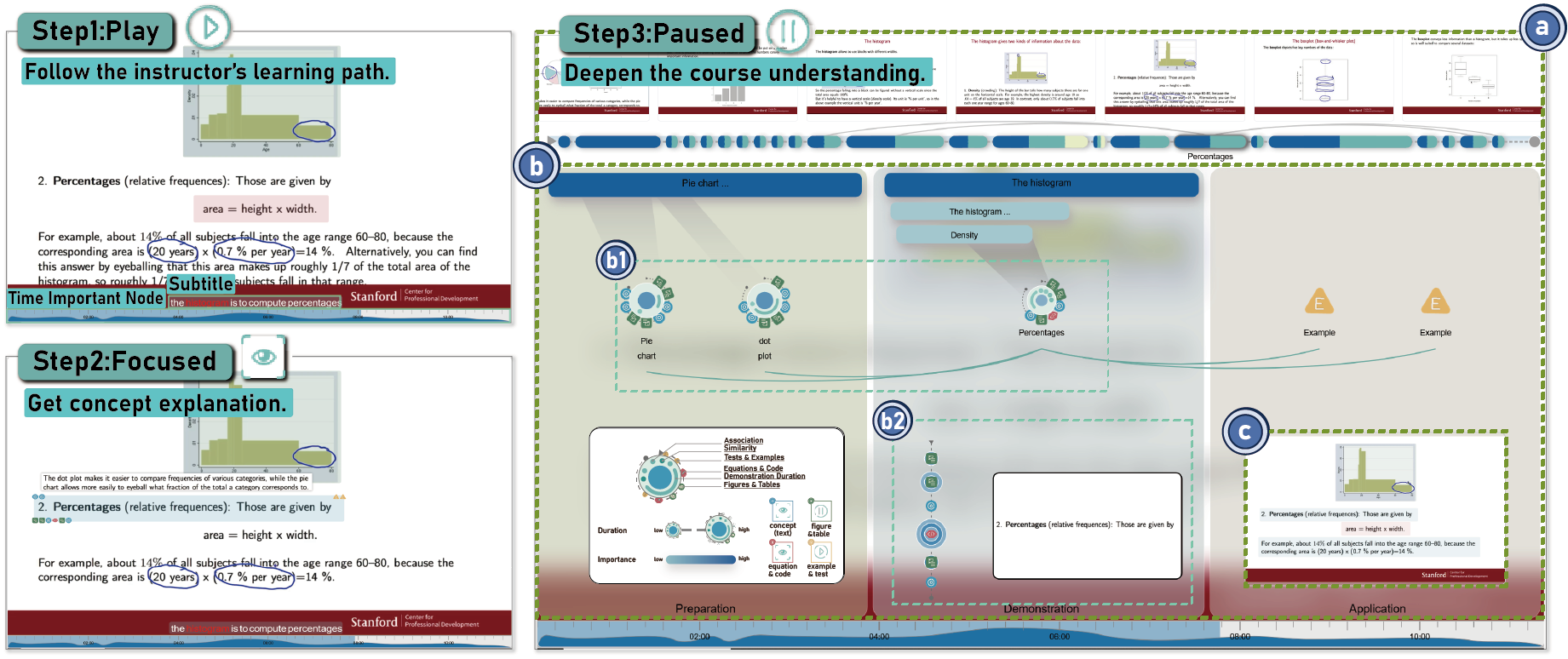} 
    \vspace{-1mm}
    \caption{\HLF{
    \toolName{} interaction workflow includes 1) \textit{Play} - Follow the instructor's learning path. 2) \textit{Focused} - Get concept explanation, and 3) \textit{Paused} - Deepen the course understanding. \textit{Paused} stage user interface, a) slides and course overview, b) video player, including (b1) concept relationships and (b2) demonstration materials, and c) video preview. Alex is currently delving into the concept of ``Percentages'' and how it relates to others.}
    }
    \vspace{-3mm}
    \label{interaction}
\end{figure*}

\vspace{-3mm}
\subsection{Usage Scenario}
\label{secIV.B}
We illustrate how \toolName{} encompasses the design requirements through the example of Alex, a hypothetical university student majoring in computer science, preparing to learn ``Fundamental Charts''\footnote{\url{https://www.coursera.org/learn/stanford-statistics/lecture/k4w4I/pie-chart-bar-graph-and-histograms}} through \toolName{} to complete his weekly tests.

\noindent
\textbf{Follow the instructor's learning path.}
Alex starts to follow the instructor's learning path using \toolName{}. He notices that the initial layout of \toolName{} contains only the video player, allowing him to focus solely on the video content (R5).
Meanwhile, the progress bar displays important concept time nodes (Fig.~\ref{play-focused}-a), giving Alex a preliminary view of the course's difficulty distribution.
As the instructor progresses, Alex notices that the content he is currently learning is \textit{Highlighted} by the light-colored rectangular box (Fig.~\ref{play-focused}-b), while the automatically generated subtitles similarly \textit{Highlight} the current concept (Fig.~\ref{play-focused}-c), which allows him to better follow the instructor's preset learning path to complete the initial learning (R1).

\noindent
\textbf{Get concept explanation.}
Currently, Alex is learning about the ``histogram'', and he finds the concept of ``percentages (relative frequencies)'' somewhat challenging to grasp.  \textcolor{black}{To facilitate his understanding, the system allows him to hover the mouse over the text to retrieve the concept.}
He finds the tool further enhances the \textit{Highlight} effect, while all elements explaining the concept are displayed around the concept with icons (Fig.~\ref{play-focused}-d, R2).
\HL{He can hover over these icons to view the original content of the elements and click to navigate.}
The same approach applies to other elements (e.g., figures, equations, etc.), with displayed content changing to the concepts contained within the elements. In this way, Alex can access concept-related material in real time to facilitate understanding.

\noindent
\textbf{Deepen the course understanding.}
After learning the ``percentages'', Alex aims to understand its relationship with other charts by ``Pause'' to enter the full mode of \toolName{}, where the original slides and concept overview are displayed at the top (Fig.~\ref{interaction}-a, R1), and the detailed cognitive path and relationships of the concepts are displayed in the video player (Fig.~\ref{interaction}-b), along with a video preview for selecting specific augmented elements (Fig.~\ref{interaction}-c).
According to the 4-MAT learning model~\cite{AboutLearning}, concept cognition is composed of \textit{Preparation}, \textit{Demonstration}, and \textit{Application}.
Therefore, the video player is divided into three areas to display content from different stages (Fig.~\ref{interaction}-b).
Alex notices that in the \textit{Preparation} of ``histogram,'' ``pie charts'' and ``dot plots'' are presented as prerequisite knowledge (Fig.~\ref{interaction}-b1, R3).
Subsequently, various elements (concepts, figures, equations) are used to \textit{Demonstrate} the content of the ``histogram'' (Fig.~\ref{interaction}-b2), culminating in their \textit{Application} in two real-world examples.
In this case, Alex gains a deeper understanding of the course's underlying logic.

Throughout the process, \toolName{} uses an intuitive and concise visual design, Alex can easily access the information (R5).
Also due to \toolName{}'s interactive support for learner behavior at all stages (R4), Alex can complete the above process once quickly, thus iteratively completing the course in the same way and better preparing for the test.

\begin{figure*}[t]
    \centering
    \vspace{-2mm}
    \includegraphics[width=\linewidth]{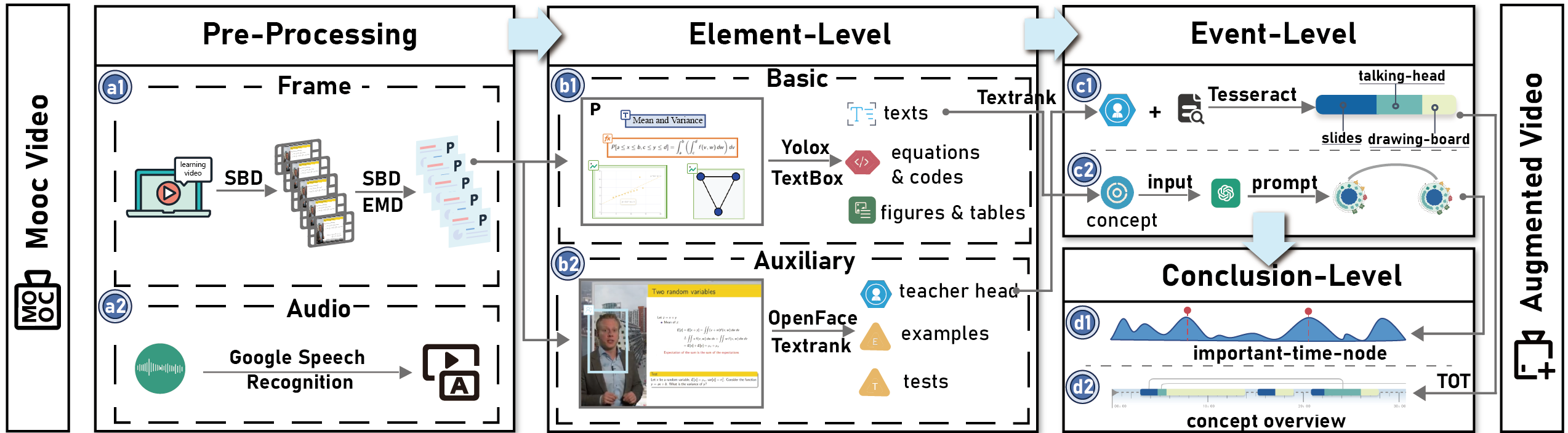}
 
    \caption{\HLF{The data processing pipeline for augmenting MOOC videos. \toolName{} adopts a bottom-up approach to (a) extract slides and audio from raw videos, then further distills these into (b) element-level, (c) event-level, and (d) conclusion-level representations through deep learning models.}}
    \vspace{-1mm}
    \label{pipeline}
\end{figure*}

\subsection{Data Processing Pipeline}
We utilize a bottom-up approach to extract data from MOOC videos.

\noindent
\textbf{Data preprocessing}. Due to the complexity of information contained in video data, we start by extracting \textit{frame} and \textit{audio} data.

\vspace{-1mm}
\noindent
\begin{itemize}[leftmargin=*]
    \setlength{\leftmargin}{0pt}
	\setlength{\itemsep}{0pt}
	\setlength{\parsep}{0pt}
	\setlength{\parskip}{0pt}

\item As for the \textit{Frame} data (Fig.~\ref{pipeline}-a1), slides are the primary objects due to their richer information. Therefore, we utilize a two-stage strategy for Slide-Based Detection (SBD)~\cite{zhao2017novel}. In the first stage, we segment slides into frames and measure the dissimilarity between frames using the Earth Mover's Distance (EMD)\cite{rubner2000earth}. Subsequently, we utilize the edge-based SBD method\cite{seaton2014does} to refine the slide's boundaries and thus distinguish each slide.

\item \textit{Audio} data (Fig.~\ref{pipeline}-a2) contains the context of concept information, which enables a more effective construction of relationships. 
We utilize Google's speech recognition \cite{SpeechRecognition} to transcribe speech into text, generating subtitles along with their corresponding timestamps.

\end{itemize}

\noindent
\textbf{Element-level data}. 
We integrate several advanced deep learning models with a rule-based algorithm for slide structure extraction to analyze the slide objects obtained from \textit{preprocessing}. 

\vspace{-1mm}
\noindent
\begin{itemize}[leftmargin=*]
    \setlength{\leftmargin}{0pt}
	\setlength{\itemsep}{0pt}
	\setlength{\parsep}{0pt}
	\setlength{\parskip}{0pt}

\item For the \textit{Basic} elements (Fig.~\ref{pipeline}-b1), our goal is to identify each element within every slide. To achieve this, we manually annotated three hundred MOOC video frames to train the YOLOX model\cite{ge2021yolox} and used this model to differentiate between text, figures, tables, equations, and code-block.  \textcolor{black}{We employed a stratified sampling method, selecting frames from courses across different disciplines (such as computer science, physics, economics, etc.), and ensured these frames included a variety of visual elements.} Additionally, TextBoxes\cite{liao2017textboxes} was employed for rapid text recognition in slides. 

\item For the \textit{Auxiliary} elements (Fig.~\ref{pipeline}-b2), OpenFace 2.0\cite{baltrusaitis2018openface} was utilized to recognize facial landmarks in shots, aiding in identifying the presence of a teacher's head. Furthermore, Textrank\cite{mihalcea2004textrank} was employed to extract keywords and key phrases, facilitating the differentiation between tests and examples.

\end{itemize}

\noindent
\textbf{Event-level data}.
Based on \textit{element-level} data, we categorized event-level data into \textit{attributes} and \textit{relationships}.

\vspace{-1mm}
\noindent
\begin{itemize}[leftmargin=*]
    \setlength{\leftmargin}{0pt}
	\setlength{\itemsep}{0pt}
	\setlength{\parsep}{0pt}
	\setlength{\parskip}{0pt}

\item To extract \textit{attributes}, \textcolor{black}{we combined text and audio, identifying the concepts through keyword extraction. Utilizing timestamps corresponding to subtitles, we determined the duration. To distinguish between slide and drawing-board, we employed Tesseract\cite{smith2009adapting} to differentiate handwritten from slide text. Coupled with the presence of the teacher's head, we deduced the delivery style (Fig.~\ref{pipeline}-c1).} 

\item Leveraging the Chain-of-Thought model proposed by Wei et al.\cite{wei2022chain}, we employed GPT-3.5 to identify concept \textit{relationships}. \textcolor{black}{We devised a series of prompts (Appendix B) to guide the GPT in inferring the three levels of relationships between the concepts (Fig.~\ref{pipeline}-c2).}
Following this, we input the text extracted from subtitles as prior knowledge and directed GPT-3.5 to generate data in a predefined JSON format, thereby clarifying the relationships between concepts.

\end{itemize}

\noindent
\textbf{Conclusion-level data}. To capture important time nodes within the course, we integrated the duration and relationships extracted from the \textit{element-level} data. A longer duration suggests a more complex concept, while a greater number of relationships could indicate a wider range of coverage (Fig.~\ref{pipeline}-d1). Additionally, we employed the Topics Over Time (TOT) model\cite{wang2006topics} to construct the course structure, which we then merged with the extracted delivery style to provide the overview of concepts (Fig.~\ref{pipeline}-d2).

\vspace{-2mm}
\subsection{Hyperlink-based Interaction Methods}
The interaction design of \toolName{} is derived by Hypervideo~\cite{chambel2006hypervideo}, which facilitates video navigation through embedded hyperlinks. 
Building on this idea and the multi-stage learning behavior of learners, we designed a multi-stage (\textit{Play}-\textit{Focused}-\textit{Paused}) interaction method through various
\hspace{-1mm}
\raisebox{-1mm}{\includegraphics[scale=0.52]{./icon-interactive.png}}
\hspace{-1mm}
\textit{Interactive Elements}.

\sideicon{./icon-play.png}%
\textit{\uline{Play \wl{(Real-time Augmentation).}}}
At this stage, \toolName{} \wl{provides synchronized 
visual cues that update in real-time during video playback,}
\hspace{-1mm}
\raisebox{-1.3mm}{\includegraphics[scale=0.52]{./icon-highlight.png}}
\hspace{-1mm}
\textit{Highlighting}
the concepts in the video frames (Fig.~\ref{play-focused}-b) and subtitles (Fig.~\ref{play-focused}-c). 
Learners can jump to important points in the course by clicking on important time nodes on the progress bar (Fig.~\ref{play-focused}-a), which is extremely beneficial for learners using \toolName{} for review.

\sideicon{./icon-focus.png}%
\textit{\uline{Focused \wl{(Real-time Detail-on-Demand)}.}}
 \textcolor{black}{Given the limited prevalence and high complexity of eye-tracking devices, \toolName{} currently implements a mouse-hover-based focus proxy as an alternative to eye-tracking. Specifically, when learners focus on a single element by hovering the mouse over it for more than 3 seconds, the tool prompts further augmentation of that element. It allows the system to identify the user's focus and provide additional explanatory information or highlighting around the concept they are attempting to understand.}

\sideicon{./icon-pause.png}%
\textit{\uline{Paused \wl{(On-demand Exploration)}.}} \wl{When learners \textit{Pause} the video, \toolName{} seamlessly transitions to its full exploratory mode, expanding the embedded visualizations to reveal comprehensive concept relationships and cognitive paths.}
The full mode globally supports two \textit{Keyboard-mouse} interaction modes: hovering and clicking. Learners can hover over concepts or other elements to display their original content or click to navigate.
When the hovered element is a concept, the content in the \textit{Demonstration} area changes to the explanation flow of the selected concept (Fig.~\ref{interaction}-b2), facilitating a continuous understanding of the knowledge.
Additionally, learners can select the currently focused concept using the video preview (Fig.~\ref{interaction}-c). 

\vspace{-2mm}
\subsection{Multi-glyph Embedded Visualizations}

\begin{figure}[tb]
    \hspace{3mm}
	\centering
	\includegraphics[width=\linewidth]{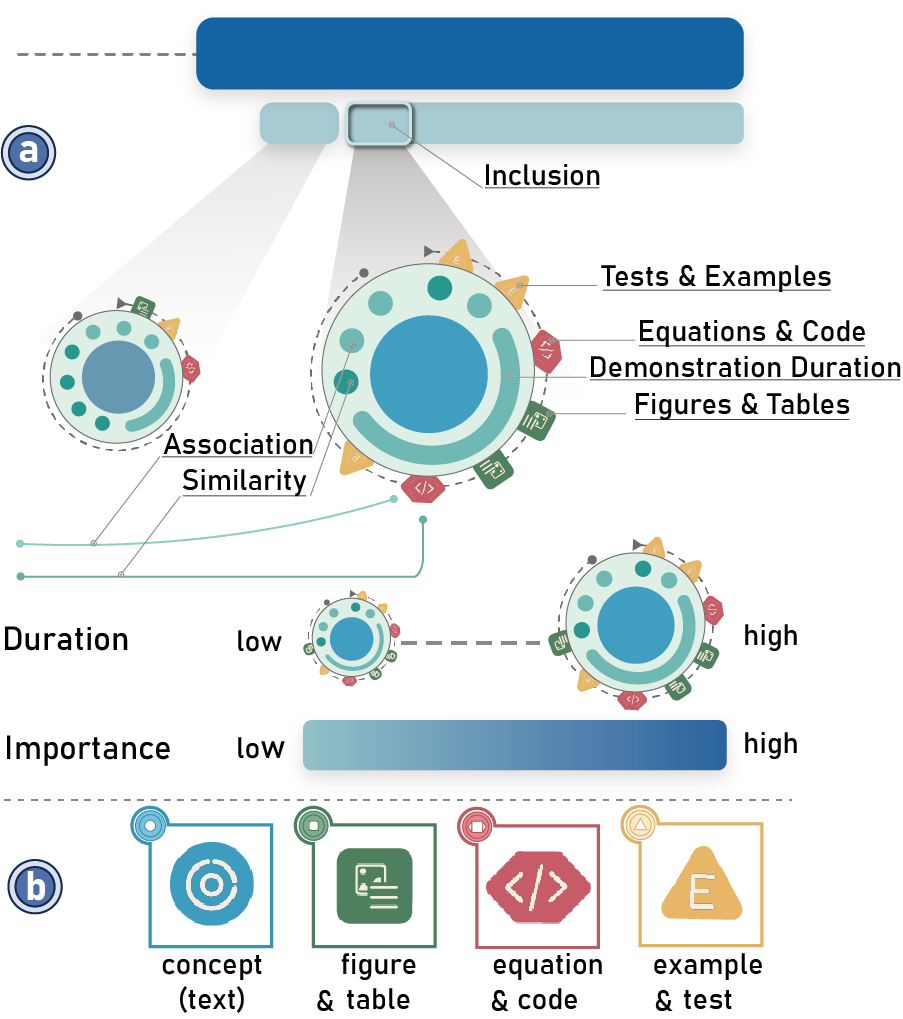}

	\caption{The multi-glyph design for basic elements. (a) shows the radial design of concepts during the \textit{Paused} stage, along with supplementary design utilized for various relationships, (b) demonstrates the glyph design of each element focusing on two key attributes: color and shape.}

	\label{visualDesign}
    \vspace{-4mm}

\end{figure}

 \textcolor{black}{Current learning content analysis systems often externalize knowledge in multiple views, requiring learners to expend additional attention on switching viewpoints. To address this, \toolName{} employs embedded visualizations.} All augmentations appear directly on the video to prevent loss of attention.

During the \textit{Play} stage, concepts being explained by the instructor are 
\hspace{-1mm}
\raisebox{-1.3mm}{\includegraphics[scale=0.52]{./icon-highlight.png}}
\hspace{-1mm}
\textit{Highlighted}
with bounding boxes corresponding to the colors of each element, and the automatically generated subtitles are similarly 
\hspace{-1mm}
\raisebox{-1.3mm}{\includegraphics[scale=0.52]{./icon-highlight.png}}
\hspace{-1mm}
\textit{Highlighted}, aiding learners in keeping up with the instructor's learning pace (Fig.~\ref{play-focused}-b).
The progress bar at the bottom displays the \textit{importance} of concepts through an overlaid area chart (Fig.~\ref{play-focused}-a).
In the \textit{Focused} stage, the color of the bounding boxes intensifies, and icons for all elements emerge around the focused object according to the concept cognition stage (Fig.~\ref{play-focused}-d).

In the \textit{Paused} stage, \toolName{} displays its full mode. Upon entering this stage, the video will shrink to the bottom through
\hspace{-1mm}
\raisebox{-1.3mm}{\includegraphics[scale=0.52]{./icon-animation.png}}
\hspace{-1mm}
\textit{Animations}.
The course's slides and \textbf{concept overview} are displayed at the top (Fig.~\ref{interaction}-a).
\textcolor{black}{Concepts are divided into up to three areas based on the \textit{delivery style}, indicating transitions in teaching methods.}
Within the video player (Fig.~\ref{interaction}-b), given that concept cognition is multi-staged~\cite{AboutLearning}, where learners first acquire prerequisite knowledge in the \textit{Preparation} stage, then understand the specific content in the \textit{Demonstration} stage, and finally apply the concept in the \textit{Application} stage, the player is divided into three areas to represent the different stages. 
The specific content is comprised of two parts: concepts and basic elements.

\begin{figure}[t]
    \hspace{0mm}
	\centering
	\includegraphics[width=\linewidth]{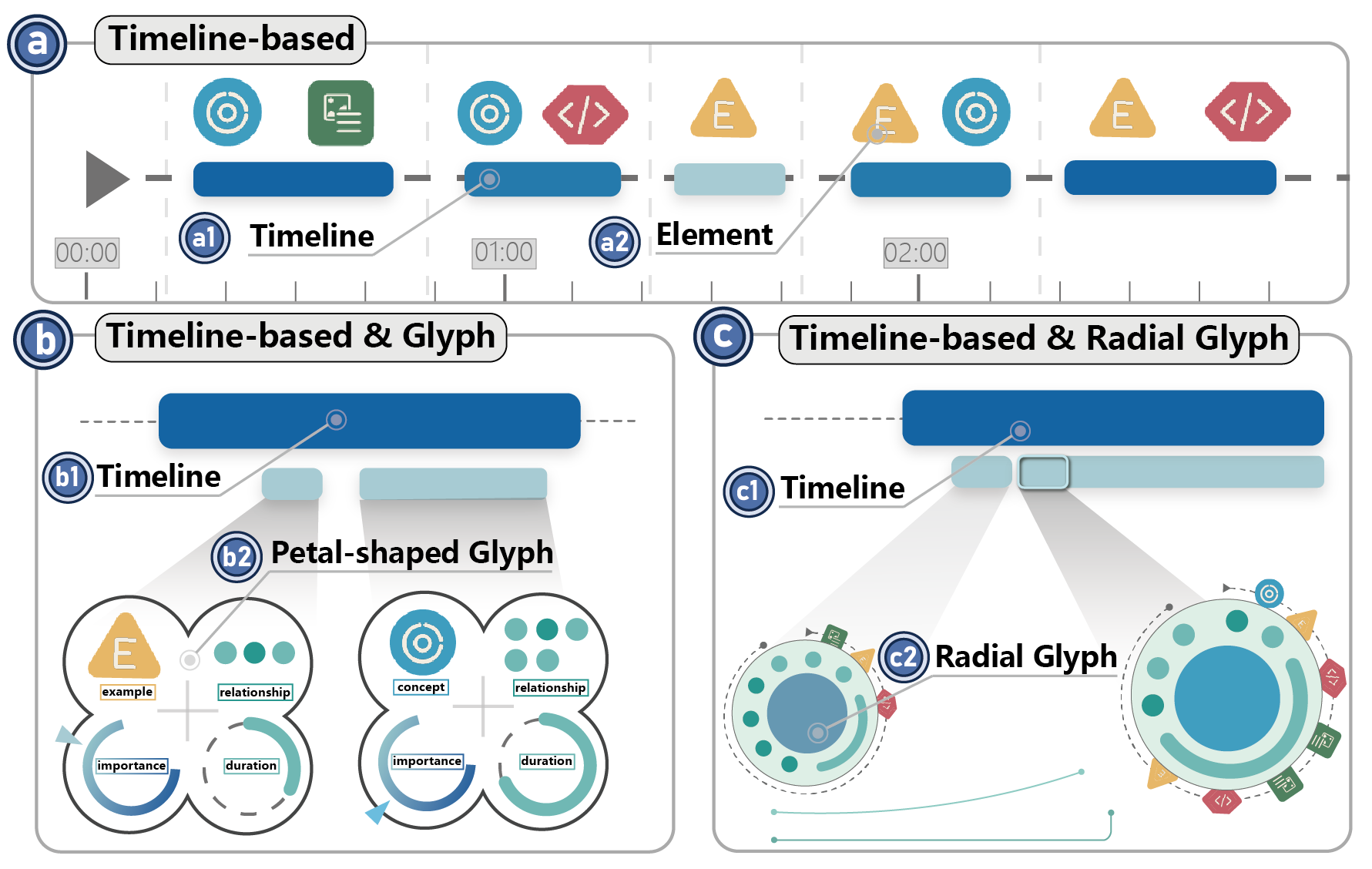}

	\caption{\textcolor{black}{The development process of our radial glyph design. The timeline-based design (a) is straightforward but suffers from clarity and usability issues as concept complexity grows. (b) To address this, we combined (b1) timelines with (b2) petal-shaped glyphs to enhance information density. (c)Finally, a (c2) radial glyph design was adopted, inspired by clock metaphors, providing a more intuitive and comprehensive visualization solution, balancing readability with rich semantic representation.}}

	\label{iteration}

\end{figure}

\textbf{Radial Glyph Design for Concepts.}
We introduce the radial design from the inside out.
\textcolor{black}{The color within a concept represents its \textit{importance} (Fig.~\ref{visualDesign}-a), with a blue gradient—from light blue indicating lower importance to dark blue indicating higher importance—determined by its duration and type of relationships..
The inner circle's ``dot-dashed line'' indicates the concept's relationships and \textit{duration} (Fig.~\ref{visualDesign}-a). Specifically, the entire inner circle maps to the course timeline. Relationships are represented by circles: 
light for \textit{associations} (and connecting curves) and dark for \textit{similarities} (and orthogonal lines).
Circular arcs depict demonstration time, with the arc size reflecting the duration. Circles and arcs are arranged in chronological order via angles. The outer ring of the radial design represents demonstration materials (Fig. \ref{visualDesign}-a), with elements coded sequentially to mirror the order in which they are mentioned. Additionally, concept size correlates with the duration (Fig. \ref{visualDesign}-a). The structures of other concepts are displayed above as layered rectangles arranged chronologically, providing a timeline view of concept evolution. The mapped lengths of these rectangles represent the durations of sub-concepts.}
 
\textbf{Multi-Glyph Design for Elements.}
We employ a series of glyph-based designs to represent all elements, encoded along two dimensions: color and shape. Specifically, concepts are represented by \textit{blue} \uline{circle}, figures \& tables by \textit{green} \uline{rectangle}, equations \& code by \textit{red} \uline{hexagon}, and examples \& tests by \textit{yellow} \uline{triangle} (Fig.~\ref{visualDesign}-b).
In the \textit{Demonstration} stage, a number of circles are applied on the outside to indicate the concepts contained within the element, with the circle colors corresponding to the color in the radial design. Also, the encoded content maintains global consistency, i.e., bounding box and float icons.

\subsection{Alternative Designs of Embedded Visualizations}\textcolor{black}{
We iterated through multiple design schemes (Fig.~\ref{iteration}) to optimize the visualization of learning concepts. Each iteration was evaluated for its fit with the system tasks:}
\textcolor{black}{\begin{enumerate}[label=(\alph*)]
\item The initial timeline-based design was simple but struggled to capture nonlinear interactions and display rich information dimensions\cite{kurihara2005flexible}.
\item We then introduced a timeline with glyphs design, utilizing multiple visual channels to describe multidimensional data attributes\cite{kovacevic2020glyph}. However, this design risked visual clutter and increased cognitive load.
\item Finally, we adopted a radial design inspired by booc.io\cite{schwab2016booc}, incorporating a clock metaphor. This design combined the linear representation of a timeline with the multi-dimensional data display advantages of a radial layout. It utilized visual channels such as shape, size, color, and direction to achieve comprehensive semantic representation while avoiding visual clutter\cite{holten2006hierarchical}. This approach improved readability while maintaining information density, effectively addressing the limitations of previous designs.
\end{enumerate}}

\textcolor{black}{The radial design was ultimately selected for its intuitive and information-rich solution for visualizing learning concepts.}

\section{Evaluation}
To evaluate \toolName{}, we conducted a user study with 36 learners and expert interviews with 2 domain experts.

\subsection{User Study}
To assess the usefulness, usability, and engagement of using hypervideo-based embedded visualizations in watching MOOC videos, we conducted a comparative study involving three learning modes based on prior evaluation methods~\cite{chen2023iball}: the raw video mode (RAW), the augmented video mode with only \textit{Play} and \textit{Focused} stages (AUG), and the full-featured version (FULL) of \toolName{}.

\noindent
\textbf{Participants and Apparatus.}
We recruited 36 participants (20 males, 16 females) via university mailing lists and forums. \HLF{This sample size was determined to provide sufficient statistical power for our between-subjects design to detect significant differences in learning outcomes and cognitive load, balancing quantitative statistical requirements with the depth of qualitative feedback required for the study. This scale aligns with contextual practices in similar augmented video research (e.g., Chen et al.~\cite{chen2023iball} used 16 participants), ensuring comparable rigor.} Participants were evenly distributed across three academic majors—Mathematics, Education, and Art Design—with 12 from each. Within each major, four participants were randomly assigned to one of three learning modes. Mathematics participants were highly familiar with the content, Education participants had partial background knowledge, and Art Design participants were encountering the material for the first time. The average age was 23.9 years (SD = 2.57). The experiment was conducted in a lab using a 24-inch monitor (1920×1080 resolution). Each session lasted approximately 70 minutes, and participants were compensated \$6.

\noindent
\textbf{Design and Procedure.}
The study followed a between-subjects design. Each participant was assigned to one learning mode (RAW, AUG, or FULL) and completed two learning tasks using that mode. Two instructional videos were selected: a scenario-based demonstration video (Fig.\ref{interaction}, Sec.\ref{secIV.B}, Task1) and a segment from “Econometrics: Random Variables”\footnote{\url{https://www.coursera.org/learn/erasmus-econometrics/lecture/2yd7H/lecture-p-1-random-variables}} (Fig.~\ref{fig:teaser}, Task2). Each was edited to a ~5-minute clip.

Participants watched both video clips in their assigned mode. For those using the AUG or FULL modes, a short tutorial (1 minute) was provided beforehand to familiarize them with interactive features. All participants had no prior exposure to these exact videos, though their subject familiarity varied.

\textit{\uline{Phase 1. Introduction (10 mins).}}
Participants were briefed on the study’s goals and procedure. After providing informed consent, they were introduced to the design concepts. For those using AUG or FULL modes, an instructional video demonstrated key features.

\textit{\uline{Phase 2. Comparative Tasks (40 mins).}}
Participants viewed the two video clips using only their assigned mode. After each clip, they completed a 10-question single-choice quiz to assess both factual recall and conceptual understanding. Participants were encouraged to focus and verbalize their thoughts while watching.

\textit{\uline{Phase 3. Post-study Questionnaire (20 mins).}}
After completing the learning tasks and quizzes, all participants were provided with brief demonstrations of the two learning modes they had not previously experienced. Each demonstration used a short (3 minutes) sample video to introduce key features of the other modes. With this full context, participants completed the System Usability Scale (SUS)\cite{brooke2013sus} and NASA-TLX~\cite{hart2006nasa} questionnaires to compare the three modes in terms of usability, cognitive load, and perceived learning support. Participants then ranked the three modes and participated in a short individual interview to share feedback.

\noindent
 \textcolor{black}{\textbf{Measures.}
We collected the following quantitative data: learning outcomes, measured by the scores of the 10 single-choice questions following each video segment (10 points per question, total 100 points); usability and cognitive load, assessed through SUS scores and NASA-TLX scores for each mode (RAW, AUG, FULL). We combined the NASA-TLX scores of the three modes, with the top third of the scores being classified as high cognitive load and the bottom third as low cognitive load. Additionally, we collected and summarized participants' feedback on the usefulness, engagement, and usability of embedded visualizations and hyperlink-based interactions in \toolName{} through interviews.}

\noindent
\textbf{Results.}
 \textcolor{black}{We first report the learning outcomes of students in the three participant groups, as well as the overall tool ratings and the cognitive rankings of the three modes,} and then discuss the feedback on the usefulness, engagement, and usability of various visualizations and interactions.
\begin{figure}[h]
	\centering
	\includegraphics[width=\linewidth]{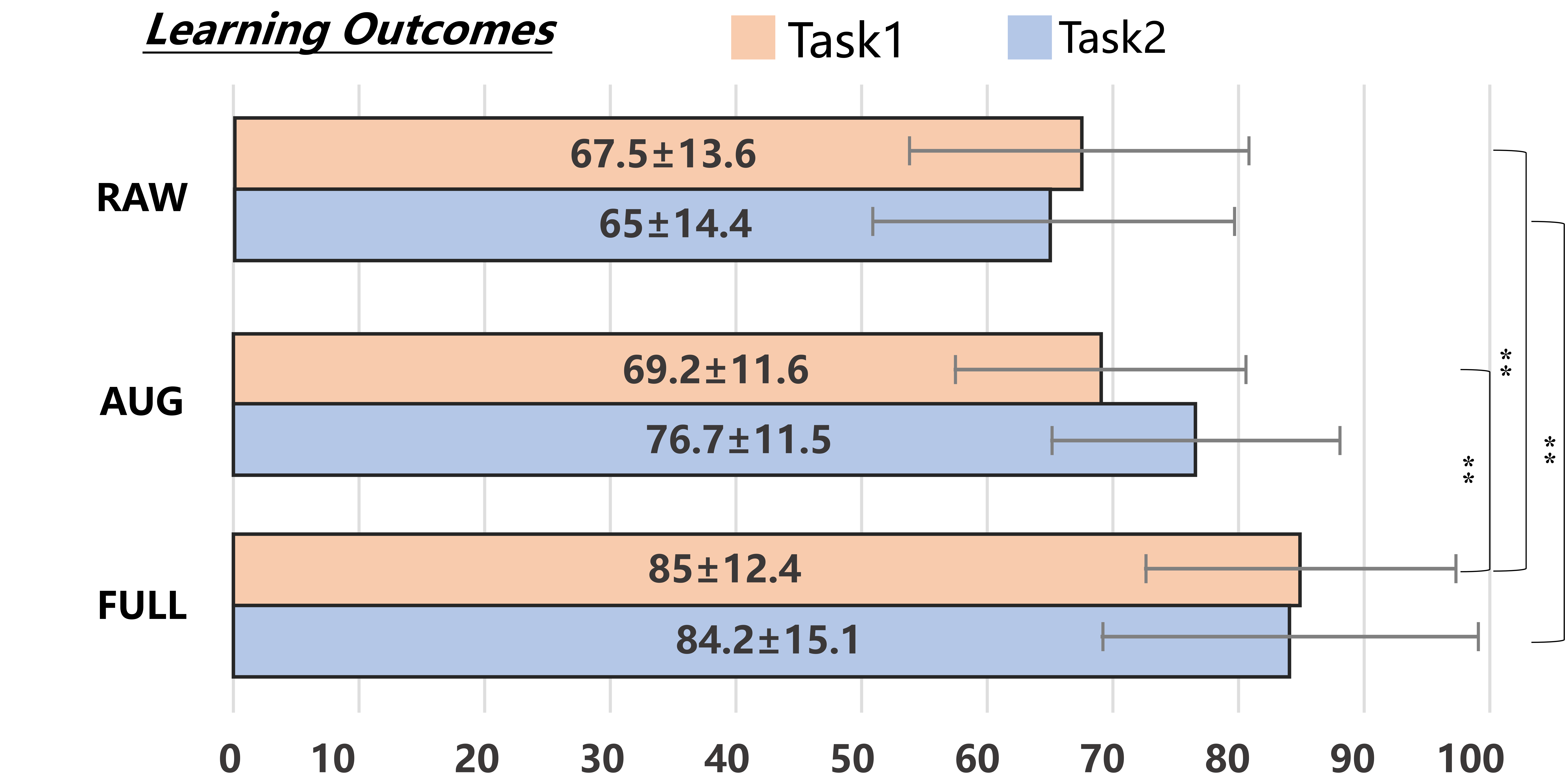}    

	\caption{\HLF{Learning test results of three participant groups across three learning modes in two tasks (the error bars indicate the standard deviation (SD) of the mean scores across participants). }}
    \vspace{-5mm}
	\label{learning}

\end{figure}    `

\begin{figure*}[htbp]
    \centering
    \includegraphics[width=\linewidth]{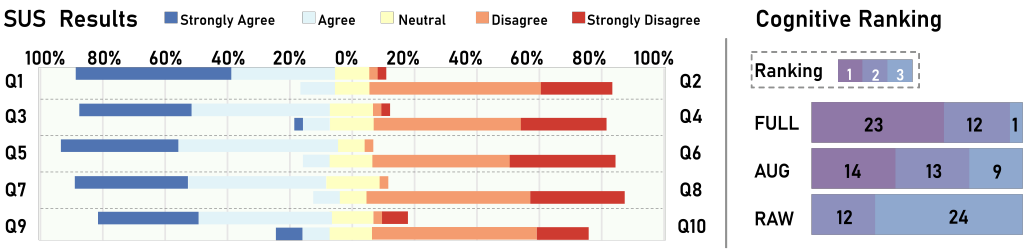}
 
    \caption{\HLF{Left: The results of the SUS. Right: Cognitive Rankings of different modes, i.e., RAW – watch with the original video, AUG – watch with \textit{Play} \& \textit{Focused} stage of \toolName{} solely, FULL – watch with the full version of \toolName{}. (For details on the SUS questionnaire items (Q1–Q10), please refer to Appendix C.)}}
    \label{user1}
\end{figure*}

\uline{\textit{Learning Outcomes.}}
\wl{One-way ANOVA revealed significant mode effects for Task 1 (F(2, 33) = 7.08, p = 0.003, $\eta^2$ = 0.30) and Task 2 (F(2, 33) = 5.90, p = 0.006, $\eta^2$ = 0.26) . Bonferroni-corrected comparisons (APPENDIX F-Table 1) showed FULL mode significantly outperformed RAW in both tasks (Task 1: p=0.010, d=1.34; Task 2: p=0.013, d=1.30), with significance markers shown in Fig.~\ref{learning}. FULL also exceeded AUG in Task 1 (p=0.012, d=1.31). Exploratory analysis revealed Education majors showed greatest improvement with FULL mode (moderate prior knowledge enabled them to leverage structured visualizations), followed by Art Design (benefited from enhanced visual support despite lower baseline) and Mathematics majors (smaller gains due to existing strong knowledge).}

\uline{\textit{Overall User Experience.}}
\wl{The results of the SUS  (Fig.~\ref{user1}  left) indicate that the majority of participants felt at ease using \toolName{} and showed a preference for utilizing it in their MOOC video learning. Fig.~\ref{user1} right shows the ranking of the three modes in terms of reducing cognitive load. Twenty-three participants (63.9\%) believed that FULL mode could significantly reduce cognitive load, followed by AUG mode with fourteen participants (38.9\%). The remaining thirteen participants (36.1\%) did not list FULL as their top choice, mainly expressing concerns that the embedded visualizations provided by \toolName{} might increase learners' cognitive load.}

User experiences varied significantly based on academic backgrounds. \toolName{} was effective for students with relevant foundational knowledge but challenging for students lacking prior knowledge. One Art Design student reported significant difficulties using the system. Most users adapted well to embedded visualization features, but a minority, mainly from Art Design, expressed concerns about advanced features potentially increasing cognitive load.

\wl{To understand these differences, we analyzed NASA-TLX scores by dimension and participant major. FULL mode (M=33.8, SD=6.7) was rated significantly lower than RAW (M=54.9, SD=6.3), indicating reduced cognitive effort in comprehension. However, this benefit varied by background: Mathematics majors rated FULL mode lowest on Mental Demand (M=29.2), while Art Design majors rated it highest (M=46.7), a 17.5-point difference (p < 0.001). Post-hoc interviews revealed that Art Design participants found the radial visualization initially ``overwhelming" (P21), whereas Mathematics participants appreciated being able to ``see the whole structure at once" (P5). This suggests that domain familiarity moderates the effectiveness of complex embedded visualizations.}
 
These findings emphasize the importance of considering user background diversity in online learning system design, especially for interdisciplinary content. They provide direction for optimizing \toolName{} and highlight the need to support learners with different knowledge backgrounds through flexible knowledge presentations or personalized learning pathways.

\begin{figure*}[htbp]
	\centering
 
	\includegraphics[width=\linewidth]{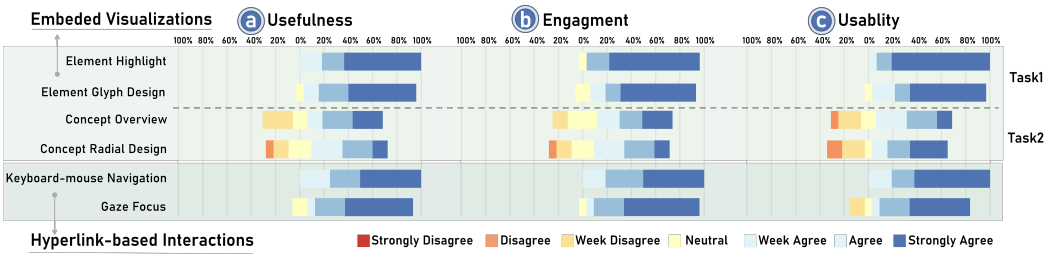}

	\caption{\HLF{Ratings on the usefulness, engagement, and usability of embedded visualizations and hyperlink-based interactions in \toolName{}. The ratings are based on post-usage user interviews.}}
	\label{user2}

\end{figure*}

\uline{\textit{Usefulness of Embedded Visualizations.}}
Participants positively evaluated the usefulness of each embedded visualization (Fig.~\ref{user2}-a).
Element Highlight was deemed most useful for following the instructor's guidance, ``\textit{I can quickly understand what the teacher is talking about}'' (P2).
Element Glyph Design was also considered useful, with participants generally agreeing that explicit indicators of concept-related elements could help them more quickly make connections between knowledge.
 \textcolor{black}{Participants had mixed opinions about the Concept Overview and the Concept Radial Design.} Most felt that pausing allowed for a deeper exploration of the concept's demonstration and connections.
However, those from art design majors (P20, P21, P22), found it increased their cognitive load. They believed that while the demo videos performed well, the relationships in their usual MOOC learning typically were not complex, One remarked, ``\textit{I prefer using the AUG mode for my studies}'' (P21).

\uline{\textit{Usefulness of hyperlink-based interactions.}}
Keyboard-mouse Navigation received positive ratings of usefulness (Fig.~\ref{user2}-a). All participants preferred using Keyboard-mouse Navigation because it helped them to immediately ``\textit{jump to the exact time point they wanted}'' (P7). 
Mouse Focus was appreciated for reducing the learning burden, as it ``\textit{directly helped me access elements of interest}'' through simple and effective interaction (P2).

\uline{\textit{Engagement of Embedded Visualizations.}}
Participants assessed the embedded visualizations as engaging (Fig.~\ref{user2}-b), and agreed that the Element Highlight and Element Glyph Design helped them to ``\textit{follow and understand the course in depth}''. As exemplified by P4, ``\textit{with this tool, one watch does what several replays would do.}''
Concept Overview and Concept Radial Design also received positive feedback. As P14 mentioned, ``\textit{I prefer to organize concepts through notes.}'' With the visualizations, he could quickly grasp the relationships between concepts without notes and thus ``\textit{facilitating review.}'' 
Several participants who strongly agreed (P1, P2, P8) felt that a comprehensive demonstration of all the elements gave them great satisfaction.

\uline{\textit{Engagement of hyperlink-based interactions.}}
The interactions provided by \toolName{} were also highly engaging (Fig.~\ref{user2}-b), as they enabled participants to  \textcolor{black}{engage in MOOC learning for the first time actively}.
P20 felt that the Mouse Focus made her feel ``\textit{in the classroom.}''
P4 agreed, stating it allowed him to interact with the instructor ``\textit{face-to-face.}''
Moreover, participants found the keyboard-mouse Navigation to be engaging, through hyperlink-based navigation, they could ``\textit{jump more precisely to the relevant time points}'' (P1), without being ``cut off'' from the video.
Overall, participants' feedback provided a strong indication that adding appropriate interactions to the learning material in the MOOC video would enable them to engage more, thereby enhancing course participation.

\uline{\textit{Overall Usability.}}
All participants confirmed the usability of embedded visualizations and hyperlink-based interactions in \toolName{} (Fig.~\ref{user2}-c).
They found both visualizations and interactions to be ``\textit{easy to understand}'' and ``\textit{use}''.
We note that some participants expressed negative opinions about the comprehensibility of the visualizations in the \textit{Paused} stage, but agreed that it was useful and just had a ``barrier to entry''.
 \textcolor{black}{In addition, some participants found the mouse-hover-based focus proxy difficult to understand, primarily because the system defaults to using a 3-second mouse hover to substitute for mouse-hover-based focus proxy.}
They suggested there should be an option allowing learners to dynamically adjust the duration to suit individual needs.

\subsection{Expert Interviews}

We additionally invited two experts (E3 \& E4) to discuss the usability and effectiveness of \toolName{}.
E3 is an educational theory researcher interested in the impact of course arrangement and organization on learners in online education, and E4 is a professor of educational psychology researching how to enhance learner engagement in online learning.
The interview began with sharing the results of the user study.
Different from the user study, we focused more on collecting feedback on the real-world use of \toolName{}.
We sought insights into how \toolName{} facilitates learning from the instructor's perspective and discussed its potential shortcomings and future improvement.

\textit{\uline{Balance active learning and cognitive load in visualizations.}
\textcolor{black}{E4 noted that while \toolName{}’s embedded visualizations provide valuable assistance, some learners reported feeling overwhelmed by the volume of information presented on the interface, which is probably due to the high information density and their limited proficiency with the proposed method at the beginning. This aligns with our user study findings. E4 mentioned that the concept map in \toolName{} effectively encourages active learning, but could benefit from user-controlled detail levels. Regarding MOOC embedded visualizations in general, E4 added, ``\textbf{Well-designed embedded visualizations with appropriate information and simplicity} can encourage question-asking and answer-seeking, especially for long-term memory tasks.''
E3 observed that learners might sometimes actively ignore visualizations, possibly due to Inattentional Blindness~\cite{healey2011attention}. To address this, E3 suggested that \textbf{embedded visualizations could incorporate multi-dimension designs to guide attention}, such as highlighting key concepts and cueing relationships.}}

\textit{\uline{Adapt visualizations to course types and difficulties.}}
E3 suggested that the design of \textbf{embedded visualizations should better meet learners' needs for explanatory and exploratory purposes} based on the course's type and difficulty. Different fields and difficulties require various types of visualizations. For instance, ``\textit{In \toolName{}, courses in literature or art, which may be relatively abstract and flat, need more explanatory visualizations to clarify content, while courses in science or engineering that have more relationships might focus on exploratory purposes,}'' E3 emphasized the importance of enabling instructors to modify visualizations within \toolName{}: ``\textit{instructors need to adjust visualizations based on the course content.}''  \textcolor{black}{This adaptability is crucial for effectively addressing the diverse needs of different academic fields and learning objectives.}

\textit{\uline{Enhance customization of hyperlink-based interactions.}}
According to the results of the user study, the customization of interactions is the main challenge that \toolName{} currently faces.
 \textcolor{black}{\textbf{Customizing hyperlink-based interactions is key to improving user engagement.}
E4, after examining \toolName{}'s current functionality, provided specific recommendations ``\toolName{} should add user-controlled switches for its various features." 
E4 further suggested that \toolName{} could allow users to ``customize the triggers and duration of these augmentations, tailoring the learning experience to their individual preferences."
To illustrate, E4 proposed adding a settings panel in \toolName{}'s video player: ``Users could select which visualizations appear and how they are triggered." This could include options for displaying concept maps, historical timelines, or mathematical formulas, and choosing triggers such as on hover, on click, or at specific time points in the video.}
Moreover, within \toolName{}, data should also be customizable (E3). MOOC videos contain a variety of data types, 
and learners need to be able to configure the data triggered by various interactions, which provide new opportunities for personalized learning~\cite{chen2008intelligent}.

\textit{\uline{Tailor interactions to diverse learner types.}}
Many works have analyzed the dropout rate of MOOC for different learning groups~\cite{shi2015vismooc,chen2016dropoutseer,zhang2023visual}, reflecting the specificity of different learner types for MOOC learning. 
Drawing from this, E4 suggests \textbf{designing adaptive hyperlink-based interactions in \toolName{} tailored to different user groups}. For instance, a beginner's initial ``pause'' might simply aim to rewatch a slide, whereas a reviewer might be more interested in understanding how concepts are related or explained. This will be a key focus of our future work.
E3 identified another learner group observed during teaching, those who prefer watching videos at double speed and frequently seeking forward. E3 believed the tool need offer features to dynamically adjust video speed based on the information density within the course and support quick jumps to key slide frames.

\textit{\uline{Expand visual effects for enhanced augmentation.}}
\toolName{} currently utilizes Graphical Marks primarily based on \textit{Texts} and \textit{Charts}. 
E3 believed that more graphical markup could be applied to augment MOOC videos, such as using statistical \textit{Figures} to represent statistical information or incorporating drawing-board style \textit{Annotations} to increase learner engagement, thus improving the efficiency of information delivery.
Similarly, Video Effects should also be expanded, for instance, by employing \textit{Animations} to explain process-oriented concepts or using \textit{Sound Effects} to enhance the classroom immersion.
However, both E3 and E4 agreed that the key to creating effective augmented MOOC videos lies in ``\textit{how creatively these elements and effects are integrated and applied to spark learners' curiosity to explore while ensuring a smooth learning process.}''

\section{\HL{Discussion}}
\HL{We provide a comprehensive analysis of \toolName{}'s modes, discuss the significance and generalization of our findings, and outline limitations and future directions.}

\subsection{\HL{Comprehensive Analysis}}
\HL{The study evaluates \toolName{} using three quantitative measures: test scores, SUS usability, and NASA-TLX workload. Overall, the FULL mode outperforms both RAW and AUG across all metrics—learners achieved higher test scores, rated the system as more usable, and mostly reported lower cognitive load. This advantage is particularly evident for learners unfamiliar with the content, suggesting that additional structure and guidance translate into substantial learning gains for these groups. However, no single mode is universally optimal: a subset of learners preferred AUG for its simpler interface while preserving key annotations, indicating that excessive embedded information may impose additional cognitive burden for complete beginners.}

\HL{From the perspective of Cognitive Load Theory (CLT)~\cite{sweller1998cognitive,Sweller2019,MutluBayraktar2019}, traditional video designs with external panels require learners to switch between interfaces while mentally reconstructing content, increasing extraneous load~\cite{Jeuris2016}. FULL mode addresses this by embedding supporting information directly into the video frame—key concepts, highlights, and navigation links are placed near explanations, reducing task-irrelevant operations and helping learners allocate more effort to comprehension~\cite{Alpizar2020,Xie2017}. The three-stage interaction flow (Play, Focused, Paused) makes an intuitive learning rhythm explicit, mitigating overload from presenting too much information at once~\cite{Spanjers2012,Liu2024}.} \HLF{From a pedagogical perspective, embedding element highlights and concept-based segmentation cues acts as instructional scaffolding to transform passive viewing into active processing~\cite{Alpizar2020,Xie2017,Liu2024}. Adhering to the spatial contiguity principle, in-situ navigation minimizes split-attention effects and preserves conceptual continuity~\cite{MutluBayraktar2019}. This facilitates metacognitive monitoring by enabling seamless, self-directed watch--review--search cycles, redirecting cognitive resources toward constructing deeper mental models rather than mere retrieval~\cite{sweller1998cognitive,Sweller2019,Jeuris2016}.}

\HL{Expert interviews complement these quantitative findings. Both experts emphasized three strengths of \toolName{}: (1) clearer structural cues for concept relationships; (2) key prompts and navigation embedded within the video, reducing interface switching; and (3) multi-stage interaction enabling on-demand exploration rather than passive reception. They also noted a boundary: for learners with weaker backgrounds, the information density in FULL can feel overwhelming initially, suggesting that a more gradual introduction or simpler default view may be needed—consistent with the preference variability in our data.}

\subsection{\HL{Significance \& Generalization}}

\HL{\toolName{} applies principles from cognitive load theory, signaling, and segmented learning~\cite{sweller1998cognitive,Sweller2019,MutluBayraktar2019,Alpizar2020,Spanjers2012,Liu2024} to hypervideo interface design. By embedding concept prompts and navigation links directly into the video frame, \toolName{} reduces the cognitive cost of interface switching~\cite{Jeuris2016}, offering a concrete example of translating learning theory into interactive design patterns. The video is thus designed as a \emph{cognitive interface} that supports and guides learning, rather than serving merely as a passive delivery medium. From an HCI perspective, \toolName{} demonstrates how embedded hypervideo can reallocate attention in information-dense learning tasks, with video serving as an active participant in organizing users' cognitive activities.}

\HL{The design method behind \toolName{} shows how to decompose design problems along multiple dimensions (data types, visual representations, user behaviors, cognitive states) and recombine them into concrete interaction strategies, then instantiate and compare different variants (RAW, AUG, FULL)~\cite{sauli2018hypervideo,Ortiz2024}. This ``decompose–recombine–evaluate'' approach generalizes beyond MOOC videos to other information-intensive systems such as programming tutorials, professional training, data-driven storytelling, and AR/VR learning environments.}

\HLF{Based on these results, we summarize three design insights applicable to broader human-computer interaction and educational contexts beyond this system:}

\begin{itemize}
    \item \textbf{\HLF{Progressive Cognitive Scaffolding.}}
    \HLF{Information hiding goes beyond UI cleanup; it is a pedagogical strategy to manage intrinsic load. By presenting essential concepts first and details only on demand, systems create a manageable learning zone. This structure prevents overload and encourages active inquiry, acting as a scaffold to support learners' gradual independence~\cite{Spanjers2012,Liu2024}.}
    
    \item \textbf{\HLF{Adaptive Knowledge Layering.}}
    \HLF{Systems should provide varying "thickness" of information to match learner expertise. Novices need explicit guidance for schema construction, while experts benefit from concise summaries. Providing flexible information levels mitigates the \textit{expertise reversal effect}, ensuring that instructional support remains effective rather than redundant as learners progress~\cite{sweller1998cognitive,Sweller2019,MutluBayraktar2019}.}
    
    \item \textbf{\HLF{Situated Active Learning.}} 
    \HLF{Placing navigation directly within the video adheres to the spatial contiguity principle, effectively minimizing split-attention. This turns passive video viewing into an active workspace. Such interaction strengthens dual coding, helping learners integrate visual and auditory information into a unified mental model without the cost of context switching~\cite{ginns2006integrating,Jeuris2016,Ortiz2024}.}
\end{itemize}



\HL{These insights are mainly derived from courses with relatively dense concepts and clear structure. In more open-ended, discussion-driven, or inspiration-oriented content, lighter-weight augmentation may be more suitable\cite{sauli2018hypervideo}. 
\HLF{More broadly, these strategies generalize to other information-rich learning environments. In immersive learning (AR/VR), the in-situ principle mitigates spatial disorientation, while Intelligent Tutoring Systems (ITS) can leverage adaptive layering to automate support fading.}
They also motivate further work in other domains to examine how information hierarchies should be defined, how display strategies can be adapted dynamically to user characteristics, and how the idea of contextual navigation can be extended to non-video complex interfaces. 
\HLF{Ultimately, framing interface elements as cognitive supports offers a blueprint for cognitive interfaces that bridge technical scalability with pedagogical depth.}}

\subsection{\HL{Future Work \& Limitations}}

\noindent
\textbf{Assessment.}
\HLF{Our work is a first attempt to augment MOOC videos with concept-based embedded visualizations, yet presents limitations. 
First, regarding our sample ($N=36$ learners, $N=4$ experts), while it aligns with contextual practices in related work~\cite{chen2023iball} and provided sufficient statistical power, the size limits broad generalizability. Future studies should recruit larger, diverse cohorts. 
Second, using academic major as a proxy for prior knowledge may confound knowledge level with quantitative skills or visualization literacy. 
Third, our metrics focus on immediate outcomes and subjective experience, overlooking long-term retention or transfer. 
Finally, the design primarily reflects learner behaviors; future work should include instructor-focused studies to better connect pedagogical intent with system design.}

\noindent
\HL{
\textbf{Interaction modalities.}
Our current prototype uses mouse hover to approximate basic Gaze Focus behavior. \HLF{While prior literature suggests mouse movement is a strong correlate of visual attention~\cite{huang2011no}, we acknowledge that it is not a perfect substitute for precise gaze data. The lack of eye-tracking verification in the current study is a limitation, as mouse movements may occasionally decouple from actual visual focus.} With eye-tracking technologies becoming increasingly affordable~\cite{silva2019eye}, future work should incorporate richer gaze-based techniques—such as fixation, saccade, and smooth pursuit~\cite{chen2023iball,Silva2018leveraging}—to better support learner interaction with embedded visualizations. Another promising direction is combining gaze with other modalities (e.g., voice and gesture) to provide more inclusive control and investigate how different modality combinations influence usability and learning outcomes.}

\noindent
\HLF{
\textbf{Pipeline generalizability.}
The data processing pipeline (Fig.~\ref{pipeline}) adopts a modular, bottom-up architecture to facilitate generalization. Most components are domain-agnostic: SBD/EMD for slide detection, Google Speech Recognition for audio transcription, OpenFace and TextRank for auxiliary element extraction, and TOT for course structure modeling. At the event level, GPT-3.5 with customizable prompts enables adaptation to different subject domains without model retraining.
However, certain limitations exist. The YOLOX model for element detection was trained on frames across multiple disciplines, but specialized visual content (e.g., medical imaging, musical notation) may require domain-specific fine-tuning. Additionally, our pipeline is primarily validated on slide-based lecture formats; non-slide formats such as laboratory demonstrations, field recordings, or whiteboard-only lectures may require extensions for element detection. Furthermore, concept-intensive STEM courses align well with our hierarchical data representation, whereas discussion-based humanities courses with less explicit concept structures may require alternative relationship modeling approaches. Future work should evaluate pipeline performance across broader MOOC platforms, content formats, and language contexts to establish generalization boundaries.}

\noindent
\HL{
\textbf{Adaptive visualization.}
Our results show substantial variation in how learners respond to different augmentation levels in terms of information density and perceived complexity, highlighting adaptive visualization as a key future direction. This involves determining how much information and which interaction capabilities should be exposed at a given moment—including concept density, relationship richness, visual encoding complexity, and advanced interactions such as filtering and path tracing. Future work can formalize complexity tiers mapping these dimensions to clear visualization configurations, and investigate tier transitions within videos and across courses. Adaptation should also account for prior knowledge differences: novices benefit from thinner information layers and stronger guidance, while advanced learners may prefer richer representations. Future systems may combine system-driven adaptation (adjusting based on pre-test scores or performance) with learner-driven customization (choosing between basic and advanced views). Furthermore, incorporating direct cognitive load indicators and physiological signals into the adaptive loop could enable real-time complexity adjustments—reducing visible elements when load is high and offering richer visualizations when load is low. Designing and validating such cognitive-load-aware adaptive strategies constitutes a substantial research agenda spanning visualization, learning analytics, and HCI.}

\vspace{-2mm}
\section{Conclusion}
This work augments MOOC videos with embedded visualizations to promote learners' course engagement and immersion.
We derived a design space through literature review, user interviews, and expert feedback, representing the objects (learning content) and subjects (learners) of MOOC learning.
Based on this design space, we developed \toolName{}, a fast prototyping tool utilizing embedded visualizations for hypervideo, with multi-glyph designs for learning content and 
hyperlink-based interactions based on learner behavior to deepen course understanding.
With \toolName{}, learners can interactively explore course content in MOOC videos, avoiding the distraction of limited attention.
A user study with 36 real-world learners shows high satisfaction with \toolName{}, confirming that participants could deepen their course understanding rapidly while learning MOOC videos.
Interviews with two domain experts reveal \toolName{}’s limitations and provide insightful implications for future improvements and opportunities.

\begin{acks} 
This work was supported in part by the National Natural Science Foundation of China.(Nos. 62277013 and 62177040), the National Social Science Fund Major Project (No. 24\&ZD075), the Zhejiang Lingyan Project (No. 2026C02A1238), the Central Government-Guided Local Science and Technology Development Fund (Grant No. 2025ZY01045) and the Zhejiang Provincial Key Laboratory of Brain Computer Collaborative Intelligence Technology and Applications(Grant No.2025E10015).
\end{acks} 

\bibliographystyle{ACM-Reference-Format}

\begin{thebibliography}{92}


\ifx \showCODEN    \undefined \def \showCODEN     #1{\unskip}     \fi
\ifx \showISBNx    \undefined \def \showISBNx     #1{\unskip}     \fi
\ifx \showISBNxiii \undefined \def \showISBNxiii  #1{\unskip}     \fi
\ifx \showISSN     \undefined \def \showISSN      #1{\unskip}     \fi
\ifx \showLCCN     \undefined \def \showLCCN      #1{\unskip}     \fi
\ifx \shownote     \undefined \def \shownote      #1{#1}          \fi
\ifx \showarticletitle \undefined \def \showarticletitle #1{#1}   \fi
\ifx \showURL      \undefined \def \showURL       {\relax}        \fi
\providecommand\bibfield[2]{#2}
\providecommand\bibinfo[2]{#2}
\providecommand\natexlab[1]{#1}
\providecommand\showeprint[2][]{arXiv:#2}

\bibitem[Adcock et~al\mbox{.}(2010)]%
        {adcock2010talkminer}
\bibfield{author}{\bibinfo{person}{John Adcock}, \bibinfo{person}{Matthew Cooper}, \bibinfo{person}{Laurent Denoue}, \bibinfo{person}{Hamed Pirsiavash}, {and} \bibinfo{person}{Lawrence~A Rowe}.} \bibinfo{year}{2010}\natexlab{}.
\newblock \showarticletitle{Talkminer: a lecture webcast search engine}. In \bibinfo{booktitle}{\emph{Proceedings of the 18th ACM International Conference on Multimedia}}. \bibinfo{pages}{241--250}.
\newblock


\bibitem[Albahr et~al\mbox{.}(2019)]%
        {albahr2019semkeyphrase}
\bibfield{author}{\bibinfo{person}{Abdulaziz Albahr}, \bibinfo{person}{Dunren Che}, {and} \bibinfo{person}{Marwan Albahar}.} \bibinfo{year}{2019}\natexlab{}.
\newblock \showarticletitle{Semkeyphrase: An unsupervised approach to keyphrase extraction from mooc video lectures}. In \bibinfo{booktitle}{\emph{Proceedings of the IEEE/WIC/ACM International Conference on Web Intelligence}}. \bibinfo{pages}{303--307}.
\newblock


\bibitem[Alpizar et~al\mbox{.}(2020)]%
        {Alpizar2020}
\bibfield{author}{\bibinfo{person}{Diego Alpizar}, \bibinfo{person}{Olusola~O. Adesope}, {and} \bibinfo{person}{Ruth~M. Wong}.} \bibinfo{year}{2020}\natexlab{}.
\newblock \showarticletitle{A Meta-Analysis of the Signaling Principle in Multimedia Learning Environments}.
\newblock \bibinfo{journal}{\emph{Educational Technology Research and Development}} \bibinfo{volume}{68}, \bibinfo{number}{5} (\bibinfo{year}{2020}), \bibinfo{pages}{2095--2119}.
\newblock
\href{https://doi.org/10.1007/s11423-020-09760-2}{doi:\nolinkurl{10.1007/s11423-020-09760-2}}


\bibitem[Amini et~al\mbox{.}(2017)]%
        {amini2017authoring}
\bibfield{author}{\bibinfo{person}{Fereshteh Amini}, \bibinfo{person}{Nathalie~Henry Riche}, \bibinfo{person}{Bongshin Lee}, \bibinfo{person}{Andres Monroy-Hernandez}, {and} \bibinfo{person}{Pourang Irani}.} \bibinfo{year}{2017}\natexlab{}.
\newblock \showarticletitle{Authoring Data-Driven Videos with DataClips}.
\newblock \bibinfo{journal}{\emph{IEEE Transactions on Visualization and Computer Graphics}} \bibinfo{volume}{23}, \bibinfo{number}{1} (\bibinfo{year}{2017}), \bibinfo{pages}{501--510}.
\newblock
\href{https://doi.org/10.1109/TVCG.2016.2598647}{doi:\nolinkurl{10.1109/TVCG.2016.2598647}}


\bibitem[Anthony(2022)]%
        {SpeechRecognition}
\bibfield{author}{\bibinfo{person}{Zhang Anthony}.} \bibinfo{year}{2022}\natexlab{}.
\newblock \bibinfo{title}{SpeechRecognition}.
\newblock \bibinfo{howpublished}{\url{https://pypi.org/project/SpeechRecognition}}.
\newblock
\urldef\tempurl%
\url{https://pypi.org/project/SpeechRecognition/}
\showURL{%
\tempurl}


\bibitem[Ausubel et~al\mbox{.}(1978)]%
        {ausubel1978educational}
\bibfield{author}{\bibinfo{person}{David~Paul Ausubel}, \bibinfo{person}{Joseph~Donald Novak}, \bibinfo{person}{Helen Hanesian}, {et~al\mbox{.}}} \bibinfo{year}{1978}\natexlab{}.
\newblock \showarticletitle{Educational psychology: A cognitive view}.
\newblock  (\bibinfo{year}{1978}).
\newblock


\bibitem[Baltrusaitis et~al\mbox{.}(2018)]%
        {baltrusaitis2018openface}
\bibfield{author}{\bibinfo{person}{Tadas Baltrusaitis}, \bibinfo{person}{Amir Zadeh}, \bibinfo{person}{Yao~Chong Lim}, {and} \bibinfo{person}{Louis-Philippe Morency}.} \bibinfo{year}{2018}\natexlab{}.
\newblock \showarticletitle{Openface 2.0: Facial behavior analysis toolkit}. In \bibinfo{booktitle}{\emph{Proceedings of the 2018 13th IEEE International Conference on Automatic Face \& Gesture Recognition (FG 2018)}}. IEEE, \bibinfo{pages}{59--66}.
\newblock
\href{https://doi.org/10.1109/FG.2018.00019}{doi:\nolinkurl{10.1109/FG.2018.00019}}


\bibitem[Blum-Smith et~al\mbox{.}(2021)]%
        {blum2021stepping}
\bibfield{author}{\bibinfo{person}{Sarah Blum-Smith}, \bibinfo{person}{Maxwell~M Yurkofsky}, {and} \bibinfo{person}{Karen Brennan}.} \bibinfo{year}{2021}\natexlab{}.
\newblock \showarticletitle{Stepping back and stepping in: Facilitating learner-centered experiences in MOOCs}.
\newblock \bibinfo{journal}{\emph{Computers \& Education}}  \bibinfo{volume}{160} (\bibinfo{year}{2021}), \bibinfo{pages}{104042}.
\newblock


\bibitem[Brooke(2013)]%
        {brooke2013sus}
\bibfield{author}{\bibinfo{person}{John Brooke}.} \bibinfo{year}{2013}\natexlab{}.
\newblock \showarticletitle{SUS: a retrospective.}
\newblock \bibinfo{journal}{\emph{Journal of Usability Studies}} \bibinfo{volume}{8}, \bibinfo{number}{2} (\bibinfo{year}{2013}).
\newblock


\bibitem[Ca{\~n}as and Novak(2014)]%
        {canas2014concept}
\bibfield{author}{\bibinfo{person}{Alberto~J Ca{\~n}as} {and} \bibinfo{person}{Joseph~D Novak}.} \bibinfo{year}{2014}\natexlab{}.
\newblock \showarticletitle{Concept mapping using CmapTools to enhance meaningful learning}.
\newblock In \bibinfo{booktitle}{\emph{Knowledge cartography: Software tools and Mapping Techniques}}. \bibinfo{publisher}{Springer}, \bibinfo{pages}{23--45}.
\newblock


\bibitem[Chambel et~al\mbox{.}(2006)]%
        {chambel2006hypervideo}
\bibfield{author}{\bibinfo{person}{Teresa Chambel}, \bibinfo{person}{Carmen Zahn}, {and} \bibinfo{person}{Matthias Finke}.} \bibinfo{year}{2006}\natexlab{}.
\newblock \showarticletitle{Hypervideo and cognition: Designing video-based hypermedia for individual learning and collaborative knowledge building}.
\newblock In \bibinfo{booktitle}{\emph{Cognitively Informed Systems: Utilizing Practical Approaches to Enrich Information Presentation and Transfer}}. \bibinfo{publisher}{IGI Global}, \bibinfo{pages}{26--49}.
\newblock


\bibitem[Chang et~al\mbox{.}(2001)]%
        {chang2001learning}
\bibfield{author}{\bibinfo{person}{Kuo-En Chang}, \bibinfo{person}{Yao-Ting Sung}, {and} \bibinfo{person}{Sung-Fang Chen}.} \bibinfo{year}{2001}\natexlab{}.
\newblock \showarticletitle{Learning through computer-based concept mapping with scaffolding aid}.
\newblock \bibinfo{journal}{\emph{Journal of Computer Assisted Learning}} \bibinfo{volume}{17}, \bibinfo{number}{1} (\bibinfo{year}{2001}), \bibinfo{pages}{21--33}.
\newblock


\bibitem[Chen(2008)]%
        {chen2008intelligent}
\bibfield{author}{\bibinfo{person}{Chih-Ming Chen}.} \bibinfo{year}{2008}\natexlab{}.
\newblock \showarticletitle{Intelligent web-based learning system with personalized learning path guidance}.
\newblock \bibinfo{journal}{\emph{Computers \& Education}} \bibinfo{volume}{51}, \bibinfo{number}{2} (\bibinfo{year}{2008}), \bibinfo{pages}{787--814}.
\newblock


\bibitem[Chen and Wu(2015)]%
        {chen2015effects}
\bibfield{author}{\bibinfo{person}{Chih-Ming Chen} {and} \bibinfo{person}{Chung-Hsin Wu}.} \bibinfo{year}{2015}\natexlab{}.
\newblock \showarticletitle{Effects of different video lecture types on sustained attention, emotion, cognitive load, and learning performance}.
\newblock \bibinfo{journal}{\emph{Computers \& Education}}  \bibinfo{volume}{80} (\bibinfo{year}{2015}), \bibinfo{pages}{108--121}.
\newblock


\bibitem[Chen et~al\mbox{.}(2017)]%
        {chen2017using}
\bibfield{author}{\bibinfo{person}{Ouhao Chen}, \bibinfo{person}{Geoff Woolcott}, {and} \bibinfo{person}{John Sweller}.} \bibinfo{year}{2017}\natexlab{}.
\newblock \showarticletitle{Using cognitive load theory to structure computer-based learning including MOOCs}.
\newblock \bibinfo{journal}{\emph{Journal of Computer Assisted Learning}} \bibinfo{volume}{33}, \bibinfo{number}{4} (\bibinfo{year}{2017}), \bibinfo{pages}{293--305}.
\newblock


\bibitem[Chen et~al\mbox{.}(2015)]%
        {chen2015peakvizor}
\bibfield{author}{\bibinfo{person}{Qing Chen}, \bibinfo{person}{Yuanzhe Chen}, \bibinfo{person}{Dongyu Liu}, \bibinfo{person}{Conglei Shi}, \bibinfo{person}{Yingcai Wu}, {and} \bibinfo{person}{Huamin Qu}.} \bibinfo{year}{2015}\natexlab{}.
\newblock \showarticletitle{Peakvizor: Visual analytics of peaks in video clickstreams from massive open online courses}.
\newblock \bibinfo{journal}{\emph{IEEE Transactions on Visualization and Computer Graphics}} \bibinfo{volume}{22}, \bibinfo{number}{10} (\bibinfo{year}{2015}), \bibinfo{pages}{2315--2330}.
\newblock


\bibitem[Chen et~al\mbox{.}(2018)]%
        {chen2018viseq}
\bibfield{author}{\bibinfo{person}{Qing Chen}, \bibinfo{person}{Xuanwu Yue}, \bibinfo{person}{Xavier Plantaz}, \bibinfo{person}{Yuanzhe Chen}, \bibinfo{person}{Conglei Shi}, \bibinfo{person}{Ting-Chuen Pong}, {and} \bibinfo{person}{Huamin Qu}.} \bibinfo{year}{2018}\natexlab{}.
\newblock \showarticletitle{Viseq: Visual analytics of learning sequence in massive open online courses}.
\newblock \bibinfo{journal}{\emph{IEEE Transactions on Visualization and Computer Graphics}} \bibinfo{volume}{26}, \bibinfo{number}{3} (\bibinfo{year}{2018}), \bibinfo{pages}{1622--1636}.
\newblock


\bibitem[Chen et~al\mbox{.}(2016)]%
        {chen2016dropoutseer}
\bibfield{author}{\bibinfo{person}{Yuanzhe Chen}, \bibinfo{person}{Qing Chen}, \bibinfo{person}{Mingqian Zhao}, \bibinfo{person}{Sebastien Boyer}, \bibinfo{person}{Kalyan Veeramachaneni}, {and} \bibinfo{person}{Huamin Qu}.} \bibinfo{year}{2016}\natexlab{}.
\newblock \showarticletitle{DropoutSeer: Visualizing learning patterns in Massive Open Online Courses for dropout reasoning and prediction}. In \bibinfo{booktitle}{\emph{Proceedings of the 2016 IEEE Conference on Visual Analytics Science and Technology (VAST)}}. IEEE, \bibinfo{pages}{111--120}.
\newblock


\bibitem[Chen et~al\mbox{.}(2023)]%
        {chen2023iball}
\bibfield{author}{\bibinfo{person}{Zhutian Chen}, \bibinfo{person}{Qisen Yang}, \bibinfo{person}{Jiarui Shan}, \bibinfo{person}{Tica Lin}, \bibinfo{person}{Johanna Beyer}, \bibinfo{person}{Haijun Xia}, {and} \bibinfo{person}{Hanspeter Pfister}.} \bibinfo{year}{2023}\natexlab{}.
\newblock \showarticletitle{iball: Augmenting basketball videos with gaze-moderated embedded visualizations}. In \bibinfo{booktitle}{\emph{Proceedings of the 2023 CHI Conference on Human Factors in Computing Systems}}. \bibinfo{pages}{1--18}.
\newblock


\bibitem[Chen et~al\mbox{.}(2021)]%
        {chen2021augmenting}
\bibfield{author}{\bibinfo{person}{Zhutian Chen}, \bibinfo{person}{Shuainan Ye}, \bibinfo{person}{Xiangtong Chu}, \bibinfo{person}{Haijun Xia}, \bibinfo{person}{Hui Zhang}, \bibinfo{person}{Huamin Qu}, {and} \bibinfo{person}{Yingcai Wu}.} \bibinfo{year}{2021}\natexlab{}.
\newblock \showarticletitle{Augmenting sports videos with viscommentator}.
\newblock \bibinfo{journal}{\emph{IEEE Transactions on Visualization and Computer Graphics}} \bibinfo{volume}{28}, \bibinfo{number}{1} (\bibinfo{year}{2021}), \bibinfo{pages}{824--834}.
\newblock


\bibitem[Chorianopoulos and Giannakos(2013)]%
        {chorianopoulos2013usability}
\bibfield{author}{\bibinfo{person}{Konstantinos Chorianopoulos} {and} \bibinfo{person}{Michail~N Giannakos}.} \bibinfo{year}{2013}\natexlab{}.
\newblock \showarticletitle{Usability design for video lectures}. In \bibinfo{booktitle}{\emph{Proceedings of the 11th European Conference on Interactive TV and video}}. \bibinfo{pages}{163--164}.
\newblock


\bibitem[Clippers(2020)]%
        {CourtVision}
\bibfield{author}{\bibinfo{person}{Clippers}.} \bibinfo{year}{2020}\natexlab{}.
\newblock \bibinfo{title}{Court Vision}.
\newblock \bibinfo{howpublished}{\url{https://www.clipperscourtvision.com/}}.
\newblock
\urldef\tempurl%
\url{https://www.clipperscourtvision.com/}
\showURL{%
\tempurl}


\bibitem[Donald(1983)]%
        {donald1983knowledge}
\bibfield{author}{\bibinfo{person}{Janet~G Donald}.} \bibinfo{year}{1983}\natexlab{}.
\newblock \showarticletitle{Knowledge structures: Methods for exploring course content}.
\newblock \bibinfo{journal}{\emph{The Journal of Higher Education}} \bibinfo{volume}{54}, \bibinfo{number}{1} (\bibinfo{year}{1983}), \bibinfo{pages}{31--41}.
\newblock


\bibitem[Dorn et~al\mbox{.}(2015)]%
        {dorn2015piloting}
\bibfield{author}{\bibinfo{person}{Brian Dorn}, \bibinfo{person}{Larissa~B Schroeder}, {and} \bibinfo{person}{Adam Stankiewicz}.} \bibinfo{year}{2015}\natexlab{}.
\newblock \showarticletitle{Piloting TrACE: Exploring spatiotemporal anchored collaboration in asynchronous learning}. In \bibinfo{booktitle}{\emph{Proceedings of the 18th ACM Conference on Computer Supported Cooperative Work \& Social Computing}}. \bibinfo{pages}{393--403}.
\newblock


\bibitem[Emmons et~al\mbox{.}(2017)]%
        {emmons2017mooc}
\bibfield{author}{\bibinfo{person}{Scott~R Emmons}, \bibinfo{person}{Robert~P Light}, {and} \bibinfo{person}{Katy B{\"o}rner}.} \bibinfo{year}{2017}\natexlab{}.
\newblock \showarticletitle{MOOC visual analytics: Empowering students, teachers, researchers, and platform developers of massively open online courses}.
\newblock \bibinfo{journal}{\emph{Journal of the Association for Information Science and Technology}} \bibinfo{volume}{68}, \bibinfo{number}{10} (\bibinfo{year}{2017}), \bibinfo{pages}{2350--2363}.
\newblock


\bibitem[Erickson(2007)]%
        {erickson2007concept}
\bibfield{author}{\bibinfo{person}{H~Lynn Erickson}.} \bibinfo{year}{2007}\natexlab{}.
\newblock \bibinfo{booktitle}{\emph{Concept-based curriculum and instruction for the thinking classroom}}.
\newblock \bibinfo{publisher}{Corwin press}.
\newblock


\bibitem[Ge et~al\mbox{.}(2021)]%
        {ge2021yolox}
\bibfield{author}{\bibinfo{person}{Zheng Ge}, \bibinfo{person}{Songtao Liu}, \bibinfo{person}{Feng Wang}, \bibinfo{person}{Zeming Li}, {and} \bibinfo{person}{Jian Sun}.} \bibinfo{year}{2021}\natexlab{}.
\newblock \showarticletitle{Yolox: Exceeding yolo series in 2021}.
\newblock \bibinfo{journal}{\emph{arXiv preprint arXiv:2107.08430}} (\bibinfo{year}{2021}).
\newblock


\bibitem[Ginns(2006)]%
        {ginns2006integrating}
\bibfield{author}{\bibinfo{person}{Paul Ginns}.} \bibinfo{year}{2006}\natexlab{}.
\newblock \showarticletitle{Integrating information: A meta-analysis of the spatial contiguity and temporal contiguity effects}.
\newblock \bibinfo{journal}{\emph{Learning and Instruction}} \bibinfo{volume}{16}, \bibinfo{number}{6} (\bibinfo{year}{2006}), \bibinfo{pages}{511--525}.
\newblock


\bibitem[Girgensohn et~al\mbox{.}(2015)]%
        {girgensohn2015hypermeeting}
\bibfield{author}{\bibinfo{person}{Andreas Girgensohn}, \bibinfo{person}{Jennifer Marlow}, \bibinfo{person}{Frank Shipman}, {and} \bibinfo{person}{Lynn Wilcox}.} \bibinfo{year}{2015}\natexlab{}.
\newblock \showarticletitle{HyperMeeting: Supporting asynchronous meetings with hypervideo}. In \bibinfo{booktitle}{\emph{Proceedings of the 23rd ACM international conference on multimedia}}. \bibinfo{pages}{611--620}.
\newblock


\bibitem[Girgensohn et~al\mbox{.}(2011)]%
        {girgensohn2011adaptive}
\bibfield{author}{\bibinfo{person}{Andreas Girgensohn}, \bibinfo{person}{Frank Shipman}, {and} \bibinfo{person}{Lynn Wilcox}.} \bibinfo{year}{2011}\natexlab{}.
\newblock \showarticletitle{Adaptive clustering and interactive visualizations to support the selection of video clips}. In \bibinfo{booktitle}{\emph{Proceedings of the 1st ACM International Conference on Multimedia Retrieval}}. \bibinfo{pages}{1--8}.
\newblock


\bibitem[Greenhow et~al\mbox{.}(2022)]%
        {greenhow2022foundations}
\bibfield{author}{\bibinfo{person}{Christine Greenhow}, \bibinfo{person}{Charles~R Graham}, {and} \bibinfo{person}{Matthew~J Koehler}.} \bibinfo{year}{2022}\natexlab{}.
\newblock \showarticletitle{Foundations of online learning: Challenges and opportunities}.
\newblock \bibinfo{journal}{\emph{Educational Psychologist}} \bibinfo{volume}{57}, \bibinfo{number}{3} (\bibinfo{year}{2022}), \bibinfo{pages}{131--147}.
\newblock


\bibitem[Hart(2006)]%
        {hart2006nasa}
\bibfield{author}{\bibinfo{person}{Sandra~G Hart}.} \bibinfo{year}{2006}\natexlab{}.
\newblock \showarticletitle{NASA-task load index (NASA-TLX); 20 years later}. In \bibinfo{booktitle}{\emph{Proceedings of the Human Factors and Ergonomics Society annual meeting}}, Vol.~\bibinfo{volume}{50}. Sage publications Sage CA: Los Angeles, CA, \bibinfo{pages}{904--908}.
\newblock


\bibitem[Haubold(2004)]%
        {haubold2004analysis}
\bibfield{author}{\bibinfo{person}{Alexander Haubold}.} \bibinfo{year}{2004}\natexlab{}.
\newblock \showarticletitle{Analysis and visualization of index words from audio transcripts of instructional videos}. In \bibinfo{booktitle}{\emph{Proceedings of the IEEE Sixth International Symposium on Multimedia Software Engineering}}. IEEE, \bibinfo{pages}{570--573}.
\newblock


\bibitem[Haubold and Kender(2005)]%
        {haubold2005augmented}
\bibfield{author}{\bibinfo{person}{Alexander Haubold} {and} \bibinfo{person}{John~R Kender}.} \bibinfo{year}{2005}\natexlab{}.
\newblock \showarticletitle{Augmented segmentation and visualization for presentation videos}. In \bibinfo{booktitle}{\emph{Proceedings Of The 13th Annual ACM International Conference On Multimedia}}. \bibinfo{pages}{51--60}.
\newblock


\bibitem[He et~al\mbox{.}(2018)]%
        {he2018vusphere}
\bibfield{author}{\bibinfo{person}{Huan He}, \bibinfo{person}{Oinghua Zheng}, {and} \bibinfo{person}{Bo Dong}.} \bibinfo{year}{2018}\natexlab{}.
\newblock \showarticletitle{VUSphere: Visual analysis of video utilization in online distance education}. In \bibinfo{booktitle}{\emph{Proceedings of the 2018 IEEE Conference on Visual Analytics Science and Technology (VAST)}}. IEEE, \bibinfo{pages}{25--35}.
\newblock


\bibitem[He et~al\mbox{.}(2023)]%
        {he2023videopro}
\bibfield{author}{\bibinfo{person}{Jianben He}, \bibinfo{person}{Xingbo Wang}, \bibinfo{person}{Kam~Kwai Wong}, \bibinfo{person}{Xijie Huang}, \bibinfo{person}{Changjian Chen}, \bibinfo{person}{Zixin Chen}, \bibinfo{person}{Fengjie Wang}, \bibinfo{person}{Min Zhu}, {and} \bibinfo{person}{Huamin Qu}.} \bibinfo{year}{2023}\natexlab{}.
\newblock \showarticletitle{Videopro: A visual analytics approach for interactive video programming}.
\newblock \bibinfo{journal}{\emph{IEEE Transactions on Visualization and Computer Graphics}} \bibinfo{volume}{30}, \bibinfo{number}{1} (\bibinfo{year}{2023}), \bibinfo{pages}{87--97}.
\newblock


\bibitem[Healey and Enns(2011)]%
        {healey2011attention}
\bibfield{author}{\bibinfo{person}{Christopher Healey} {and} \bibinfo{person}{James Enns}.} \bibinfo{year}{2011}\natexlab{}.
\newblock \showarticletitle{Attention and visual memory in visualization and computer graphics}.
\newblock \bibinfo{journal}{\emph{IEEE Transactions on Visualization and Computer Graphics}} \bibinfo{volume}{18}, \bibinfo{number}{7} (\bibinfo{year}{2011}), \bibinfo{pages}{1170--1188}.
\newblock


\bibitem[Holten(2006)]%
        {holten2006hierarchical}
\bibfield{author}{\bibinfo{person}{Danny Holten}.} \bibinfo{year}{2006}\natexlab{}.
\newblock \showarticletitle{Hierarchical edge bundles: Visualization of adjacency relations in hierarchical data}.
\newblock \bibinfo{journal}{\emph{IEEE Transactions on visualization and computer graphics}} \bibinfo{volume}{12}, \bibinfo{number}{5} (\bibinfo{year}{2006}), \bibinfo{pages}{741--748}.
\newblock


\bibitem[Huang et~al\mbox{.}(2011)]%
        {huang2011no}
\bibfield{author}{\bibinfo{person}{Jeff Huang}, \bibinfo{person}{Ryen~W White}, {and} \bibinfo{person}{Susan Dumais}.} \bibinfo{year}{2011}\natexlab{}.
\newblock \showarticletitle{No clicks, no problem: using cursor movements to understand and improve search}. In \bibinfo{booktitle}{\emph{Proceedings of the SIGCHI conference on human factors in computing systems}}. \bibinfo{pages}{1225--1234}.
\newblock


\bibitem[H{\"u}rst et~al\mbox{.}(2004)]%
        {hurst2004advanced}
\bibfield{author}{\bibinfo{person}{Wolfgang H{\"u}rst}, \bibinfo{person}{Georg G{\"o}tz}, {and} \bibinfo{person}{Philipp Jarvers}.} \bibinfo{year}{2004}\natexlab{}.
\newblock \showarticletitle{Advanced user interfaces for dynamic video browsing}. In \bibinfo{booktitle}{\emph{Proceedings of the 12th annual ACM international conference on Multimedia}}. \bibinfo{pages}{742--743}.
\newblock


\bibitem[Jeuris and Bardram(2016)]%
        {Jeuris2016}
\bibfield{author}{\bibinfo{person}{Steven Jeuris} {and} \bibinfo{person}{Jakob~E. Bardram}.} \bibinfo{year}{2016}\natexlab{}.
\newblock \showarticletitle{Dedicated Workspaces: Faster Resumption Times and Reduced Cognitive Load in Sequential Multitasking}.
\newblock \bibinfo{journal}{\emph{Computers in Human Behavior}}  \bibinfo{volume}{62} (\bibinfo{year}{2016}), \bibinfo{pages}{404--414}.
\newblock
\href{https://doi.org/10.1016/j.chb.2016.03.077}{doi:\nolinkurl{10.1016/j.chb.2016.03.077}}


\bibitem[Kizilcec et~al\mbox{.}(2015)]%
        {kizilcec2015instructor}
\bibfield{author}{\bibinfo{person}{Ren{\'e}~F Kizilcec}, \bibinfo{person}{Jeremy~N Bailenson}, {and} \bibinfo{person}{Charles~J Gomez}.} \bibinfo{year}{2015}\natexlab{}.
\newblock \showarticletitle{The instructor’s face in video instruction: Evidence from two large-scale field studies.}
\newblock \bibinfo{journal}{\emph{Journal of Educational Psychology}} \bibinfo{volume}{107}, \bibinfo{number}{3} (\bibinfo{year}{2015}), \bibinfo{pages}{724}.
\newblock


\bibitem[Kovacevic et~al\mbox{.}(2020)]%
        {kovacevic2020glyph}
\bibfield{author}{\bibinfo{person}{Nikola Kovacevic}, \bibinfo{person}{Rafael Wampfler}, \bibinfo{person}{Barbara Solenthaler}, \bibinfo{person}{Markus Gross}, {and} \bibinfo{person}{Tobias G{\"u}nther}.} \bibinfo{year}{2020}\natexlab{}.
\newblock \showarticletitle{Glyph-based visualization of affective states}. In \bibinfo{booktitle}{\emph{EuroVis 2020-22nd EG/VGTC Conference on Visualization Norrk{\"o}ping, Sweden, May 25-29, 2020}}. Eurographics Association.
\newblock


\bibitem[Kui et~al\mbox{.}(2022)]%
        {kui2022survey}
\bibfield{author}{\bibinfo{person}{Xiaoyan Kui}, \bibinfo{person}{Naiming Liu}, \bibinfo{person}{Qiang Liu}, \bibinfo{person}{Jingwei Liu}, \bibinfo{person}{Xiaoqian Zeng}, {and} \bibinfo{person}{Chao Zhang}.} \bibinfo{year}{2022}\natexlab{}.
\newblock \showarticletitle{A survey of visual analytics techniques for online education}.
\newblock \bibinfo{journal}{\emph{Visual Informatics}} \bibinfo{volume}{6}, \bibinfo{number}{4} (\bibinfo{year}{2022}), \bibinfo{pages}{67--77}.
\newblock


\bibitem[Kurihara et~al\mbox{.}(2005)]%
        {kurihara2005flexible}
\bibfield{author}{\bibinfo{person}{Kazutaka Kurihara}, \bibinfo{person}{David Vronay}, {and} \bibinfo{person}{Takeo Igarashi}.} \bibinfo{year}{2005}\natexlab{}.
\newblock \showarticletitle{Flexible timeline user interface using constraints}. In \bibinfo{booktitle}{\emph{CHI'05 Extended Abstracts on Human Factors in Computing Systems}}. \bibinfo{pages}{1581--1584}.
\newblock


\bibitem[Lasater and Nielsen(2009)]%
        {lasater2009influence}
\bibfield{author}{\bibinfo{person}{Kathie Lasater} {and} \bibinfo{person}{Ann Nielsen}.} \bibinfo{year}{2009}\natexlab{}.
\newblock \showarticletitle{The influence of concept-based learning activities on students' clinical judgment development}.
\newblock \bibinfo{journal}{\emph{Journal of Nursing Education}} \bibinfo{volume}{48}, \bibinfo{number}{8} (\bibinfo{year}{2009}), \bibinfo{pages}{441--446}.
\newblock


\bibitem[Latham(2020)]%
        {latham2020concept}
\bibfield{author}{\bibinfo{person}{Linda Latham}.} \bibinfo{year}{2020}\natexlab{}.
\newblock \showarticletitle{Concept-based education}.
\newblock In \bibinfo{booktitle}{\emph{Student-Focused Learning: Higher Education in an Exponential Digital Era}}. \bibinfo{publisher}{Rowman \& Littlefield}, \bibinfo{pages}{1--18}.
\newblock


\bibitem[Lei et~al\mbox{.}(2015)]%
        {lei2015advancing}
\bibfield{author}{\bibinfo{person}{Chi-Un Lei}, \bibinfo{person}{Xiangyu Hou}, \bibinfo{person}{Tyrone~TO Kwok}, \bibinfo{person}{Trudi~SF Chan}, \bibinfo{person}{Jane Lee}, \bibinfo{person}{Elizabeth Oh}, \bibinfo{person}{Donn Gonda}, \bibinfo{person}{Yip-Chun~Au Yeung}, {and} \bibinfo{person}{Cherry Lai}.} \bibinfo{year}{2015}\natexlab{}.
\newblock \showarticletitle{Advancing MOOC and SPOC development via a learner decision journey analytic framework}. In \bibinfo{booktitle}{\emph{Proceedings of the 2015 IEEE International Conference on Teaching, Assessment, and Learning for Engineering (TALE)}}. IEEE, \bibinfo{pages}{149--156}.
\newblock


\bibitem[Liao et~al\mbox{.}(2017)]%
        {liao2017textboxes}
\bibfield{author}{\bibinfo{person}{Minghui Liao}, \bibinfo{person}{Baoguang Shi}, \bibinfo{person}{Xiang Bai}, \bibinfo{person}{Xinggang Wang}, {and} \bibinfo{person}{Wenyu Liu}.} \bibinfo{year}{2017}\natexlab{}.
\newblock \showarticletitle{Textboxes: A fast text detector with a single deep neural network}. In \bibinfo{booktitle}{\emph{Proceedings of the AAAI Conference on Artificial Intelligence}}, Vol.~\bibinfo{volume}{31}.
\newblock


\bibitem[Liu et~al\mbox{.}(2018)]%
        {liu2018conceptscape}
\bibfield{author}{\bibinfo{person}{Ching Liu}, \bibinfo{person}{Juho Kim}, {and} \bibinfo{person}{Hao-Chuan Wang}.} \bibinfo{year}{2018}\natexlab{}.
\newblock \showarticletitle{ConceptScape: Collaborative concept mapping for video learning}. In \bibinfo{booktitle}{\emph{Proceedings of the 2018 CHI conference on human factors in computing systems}}. \bibinfo{pages}{1--12}.
\newblock


\bibitem[Liu et~al\mbox{.}(2024)]%
        {Liu2024}
\bibfield{author}{\bibinfo{person}{Dong Liu}, \bibinfo{person}{Xiaoyu Zhang}, \bibinfo{person}{Jing An}, {and} \bibinfo{person}{Yan Wang}.} \bibinfo{year}{2024}\natexlab{}.
\newblock \showarticletitle{The Effects of Segmentation on Cognitive Load, Vocabulary Learning and Retention, and Reading Comprehension in a Multimedia Learning Environment}.
\newblock \bibinfo{journal}{\emph{BMC Psychology}} \bibinfo{volume}{12}, \bibinfo{number}{1} (\bibinfo{year}{2024}), \bibinfo{pages}{4}.
\newblock
\href{https://doi.org/10.1186/s40359-023-01631-2}{doi:\nolinkurl{10.1186/s40359-023-01631-2}}


\bibitem[Mccarthy(1996)]%
        {AboutLearning}
\bibfield{author}{\bibinfo{person}{Bernice Mccarthy}.} \bibinfo{year}{1996}\natexlab{}.
\newblock \bibinfo{booktitle}{\emph{About Learning}}.
\newblock \bibinfo{publisher}{About Learning Inc}.
\newblock
\showISBNx{0960899294}


\bibitem[Mihalcea and Tarau(2004)]%
        {mihalcea2004textrank}
\bibfield{author}{\bibinfo{person}{Rada Mihalcea} {and} \bibinfo{person}{Paul Tarau}.} \bibinfo{year}{2004}\natexlab{}.
\newblock \showarticletitle{Textrank: Bringing order into text}. In \bibinfo{booktitle}{\emph{Proceedings of the 2004 Conference on Empirical Methods in Natural Language Processing}}. \bibinfo{pages}{404--411}.
\newblock


\bibitem[Miller(2015)]%
        {miller2015using}
\bibfield{author}{\bibinfo{person}{Brian~W Miller}.} \bibinfo{year}{2015}\natexlab{}.
\newblock \showarticletitle{Using reading times and eye-movements to measure cognitive engagement}.
\newblock \bibinfo{journal}{\emph{Educational Psychologist}} \bibinfo{volume}{50}, \bibinfo{number}{1} (\bibinfo{year}{2015}), \bibinfo{pages}{31--42}.
\newblock


\bibitem[Milligan and Griffin(2016)]%
        {milligan2016understanding}
\bibfield{author}{\bibinfo{person}{Sandra~Kaye Milligan} {and} \bibinfo{person}{Patrick Griffin}.} \bibinfo{year}{2016}\natexlab{}.
\newblock \showarticletitle{Understanding learning and learning design in MOOCs: A measurement-based interpretation}.
\newblock \bibinfo{journal}{\emph{Journal of Learning Analytics}} \bibinfo{volume}{3}, \bibinfo{number}{2} (\bibinfo{year}{2016}), \bibinfo{pages}{88--115}.
\newblock


\bibitem[Moorthy et~al\mbox{.}(2020)]%
        {moorthy2020gazed}
\bibfield{author}{\bibinfo{person}{KL~Bhanu Moorthy}, \bibinfo{person}{Moneish Kumar}, \bibinfo{person}{Ramanathan Subramanian}, {and} \bibinfo{person}{Vineet Gandhi}.} \bibinfo{year}{2020}\natexlab{}.
\newblock \showarticletitle{Gazed--gaze-guided cinematic editing of wide-angle monocular video recordings}. In \bibinfo{booktitle}{\emph{Proceedings of the 2020 CHI Conference on Human Factors in Computing Systems}}. \bibinfo{pages}{1--11}.
\newblock


\bibitem[Mutlu-Bayraktar et~al\mbox{.}(2019)]%
        {MutluBayraktar2019}
\bibfield{author}{\bibinfo{person}{Derya Mutlu-Bayraktar}, \bibinfo{person}{Vildan Co{\c s}kun}, {and} \bibinfo{person}{Tuba Altan}.} \bibinfo{year}{2019}\natexlab{}.
\newblock \showarticletitle{Cognitive Load in Multimedia Learning Environments: A Systematic Review}.
\newblock \bibinfo{journal}{\emph{Computers \& Education}}  \bibinfo{volume}{141} (\bibinfo{year}{2019}), \bibinfo{pages}{103618}.
\newblock
\href{https://doi.org/10.1016/j.compedu.2019.103618}{doi:\nolinkurl{10.1016/j.compedu.2019.103618}}


\bibitem[Oliver(2009)]%
        {oliver2009investigation}
\bibfield{author}{\bibinfo{person}{Kevin Oliver}.} \bibinfo{year}{2009}\natexlab{}.
\newblock \showarticletitle{An investigation of concept mapping to improve the reading comprehension of science texts}.
\newblock \bibinfo{journal}{\emph{Journal of Science Education and Technology}}  \bibinfo{volume}{18} (\bibinfo{year}{2009}), \bibinfo{pages}{402--414}.
\newblock


\bibitem[Ortiz and Moya(2024)]%
        {Ortiz2024}
\bibfield{author}{\bibinfo{person}{Mar{\'i}a~Jos{\'e} Ortiz} {and} \bibinfo{person}{Jos{\'e}~Antonio Moya}.} \bibinfo{year}{2024}\natexlab{}.
\newblock \showarticletitle{Hypervideo as a Tool for Interactive Advertising}.
\newblock \bibinfo{journal}{\emph{Communication \& Society}} \bibinfo{volume}{37}, \bibinfo{number}{1} (\bibinfo{year}{2024}), \bibinfo{pages}{21--40}.
\newblock
\href{https://doi.org/10.15581/003.37.1.21-40}{doi:\nolinkurl{10.15581/003.37.1.21-40}}


\bibitem[Ozan and Ozarslan(2016)]%
        {ozan2016video}
\bibfield{author}{\bibinfo{person}{Ozlem Ozan} {and} \bibinfo{person}{Yasin Ozarslan}.} \bibinfo{year}{2016}\natexlab{}.
\newblock \showarticletitle{Video lecture watching behaviors of learners in online courses}.
\newblock \bibinfo{journal}{\emph{Educational Media International}} \bibinfo{volume}{53}, \bibinfo{number}{1} (\bibinfo{year}{2016}), \bibinfo{pages}{27--41}.
\newblock


\bibitem[Pilli and Admiraal(2016)]%
        {pilli2016taxonomy}
\bibfield{author}{\bibinfo{person}{Olga Pilli} {and} \bibinfo{person}{Wilfried Admiraal}.} \bibinfo{year}{2016}\natexlab{}.
\newblock \showarticletitle{A taxonomy of massive open online courses}.
\newblock \bibinfo{journal}{\emph{Contemporary Educational Technology}} \bibinfo{volume}{7}, \bibinfo{number}{3} (\bibinfo{year}{2016}), \bibinfo{pages}{223--240}.
\newblock


\bibitem[Rachavarapu et~al\mbox{.}(2018)]%
        {rachavarapu2018watch}
\bibfield{author}{\bibinfo{person}{Kranthi~Kumar Rachavarapu}, \bibinfo{person}{Moneish Kumar}, \bibinfo{person}{Vineet Gandhi}, {and} \bibinfo{person}{Ramanathan Subramanian}.} \bibinfo{year}{2018}\natexlab{}.
\newblock \showarticletitle{Watch to edit: Video retargeting using gaze}. In \bibinfo{booktitle}{\emph{Computer Graphics Forum}}, Vol.~\bibinfo{volume}{37}. Wiley Online Library, \bibinfo{pages}{205--215}.
\newblock


\bibitem[Rahman et~al\mbox{.}(2023)]%
        {rahman2023enhancing}
\bibfield{author}{\bibinfo{person}{Mohammad~Rajiur Rahman}, \bibinfo{person}{Raga~Shalini Koka}, \bibinfo{person}{Shishir~K Shah}, \bibinfo{person}{Thamar Solorio}, {and} \bibinfo{person}{Jaspal Subhlok}.} \bibinfo{year}{2023}\natexlab{}.
\newblock \showarticletitle{Enhancing lecture video navigation with AI generated summaries}.
\newblock \bibinfo{journal}{\emph{Education and Information Technologies}}  \bibinfo{volume}{1} (\bibinfo{year}{2023}), \bibinfo{pages}{1--24}.
\newblock


\bibitem[Rubner et~al\mbox{.}(2000)]%
        {rubner2000earth}
\bibfield{author}{\bibinfo{person}{Yossi Rubner}, \bibinfo{person}{Carlo Tomasi}, {and} \bibinfo{person}{Leonidas~J Guibas}.} \bibinfo{year}{2000}\natexlab{}.
\newblock \showarticletitle{The earth mover's distance as a metric for image retrieval}.
\newblock \bibinfo{journal}{\emph{International Journal of Computer Vision}} \bibinfo{volume}{40}, \bibinfo{number}{2} (\bibinfo{year}{2000}), \bibinfo{pages}{99}.
\newblock
\href{https://doi.org/10.1023/A:1026543900054}{doi:\nolinkurl{10.1023/A:1026543900054}}


\bibitem[Sacha et~al\mbox{.}(2016)]%
        {sacha2016human}
\bibfield{author}{\bibinfo{person}{Dominik Sacha}, \bibinfo{person}{Michael Sedlmair}, \bibinfo{person}{Leishi Zhang}, \bibinfo{person}{J Lee}, \bibinfo{person}{Daniel Weiskopf}, \bibinfo{person}{Stephen North}, {and} \bibinfo{person}{Daniel Keim}.} \bibinfo{year}{2016}\natexlab{}.
\newblock \showarticletitle{Human-centered machine learning through interactive visualization}. In \bibinfo{booktitle}{\emph{Proceedings of the 24th European Symposium on Artificial Neural Networks, Computational Intelligence and Machine Learning}}. 85q38, \bibinfo{pages}{641--646}.
\newblock


\bibitem[Sauli et~al\mbox{.}(2018)]%
        {sauli2018hypervideo}
\bibfield{author}{\bibinfo{person}{Florinda Sauli}, \bibinfo{person}{Alberto Cattaneo}, {and} \bibinfo{person}{Hans van~der Meij}.} \bibinfo{year}{2018}\natexlab{}.
\newblock \showarticletitle{Hypervideo for educational purposes: a literature review on a multifaceted technological tool}.
\newblock \bibinfo{journal}{\emph{Technology, Pedagogy and Education}} \bibinfo{volume}{27}, \bibinfo{number}{1} (\bibinfo{year}{2018}), \bibinfo{pages}{115--134}.
\newblock


\bibitem[Schwab et~al\mbox{.}(2016)]%
        {schwab2016booc}
\bibfield{author}{\bibinfo{person}{Michail Schwab}, \bibinfo{person}{Hendrik Strobelt}, \bibinfo{person}{James Tompkin}, \bibinfo{person}{Colin Fredericks}, \bibinfo{person}{Connor Huff}, \bibinfo{person}{Dana Higgins}, \bibinfo{person}{Anton Strezhnev}, \bibinfo{person}{Mayya Komisarchik}, \bibinfo{person}{Gary King}, {and} \bibinfo{person}{Hanspeter Pfister}.} \bibinfo{year}{2016}\natexlab{}.
\newblock \showarticletitle{booc. io: An education system with hierarchical concept maps and dynamic non-linear learning plans}.
\newblock \bibinfo{journal}{\emph{IEEE Transactions on Visualization and Computer Graphics}} \bibinfo{volume}{23}, \bibinfo{number}{1} (\bibinfo{year}{2016}), \bibinfo{pages}{571--580}.
\newblock


\bibitem[Seaton et~al\mbox{.}(2014)]%
        {seaton2014does}
\bibfield{author}{\bibinfo{person}{Daniel~T Seaton}, \bibinfo{person}{Yoav Bergner}, \bibinfo{person}{Isaac Chuang}, \bibinfo{person}{Piotr Mitros}, {and} \bibinfo{person}{David~E Pritchard}.} \bibinfo{year}{2014}\natexlab{}.
\newblock \showarticletitle{Who does what in a massive open online course?}
\newblock \bibinfo{journal}{\emph{Commun. ACM}} \bibinfo{volume}{57}, \bibinfo{number}{4} (\bibinfo{year}{2014}), \bibinfo{pages}{58--65}.
\newblock
\href{https://doi.org/10.1145/2500876}{doi:\nolinkurl{10.1145/2500876}}


\bibitem[Shi et~al\mbox{.}(2015)]%
        {shi2015vismooc}
\bibfield{author}{\bibinfo{person}{Conglei Shi}, \bibinfo{person}{Siwei Fu}, \bibinfo{person}{Qing Chen}, {and} \bibinfo{person}{Huamin Qu}.} \bibinfo{year}{2015}\natexlab{}.
\newblock \showarticletitle{VisMOOC: Visualizing video clickstream data from massive open online courses}. In \bibinfo{booktitle}{\emph{Proceedings of the 2015 IEEE Pacific Visualization Symposium (PacificVis)}}. IEEE, \bibinfo{pages}{159--166}.
\newblock


\bibitem[Shih and Huang(2005)]%
        {shih2005content}
\bibfield{author}{\bibinfo{person}{Huang-Chia Shih} {and} \bibinfo{person}{Chung-Lin Huang}.} \bibinfo{year}{2005}\natexlab{}.
\newblock \showarticletitle{Content-based multi-functional video retrieval system}. In \bibinfo{booktitle}{\emph{Proceedings of the 2005 Digest of Technical Papers. International Conference on Consumer Electronics, 2005.}} IEEE, \bibinfo{pages}{383--384}.
\newblock


\bibitem[Silva et~al\mbox{.}(2019)]%
        {silva2019eye}
\bibfield{author}{\bibinfo{person}{Nelson Silva}, \bibinfo{person}{Tanja Blascheck}, \bibinfo{person}{Radu Jianu}, \bibinfo{person}{Nils Rodrigues}, \bibinfo{person}{Daniel Weiskopf}, \bibinfo{person}{Martin Raubal}, {and} \bibinfo{person}{Tobias Schreck}.} \bibinfo{year}{2019}\natexlab{}.
\newblock \showarticletitle{Eye tracking support for visual analytics systems: foundations, current applications, and research challenges}. In \bibinfo{booktitle}{\emph{Proceedings of the 11th ACM Symposium on Eye Tracking Research \& Applications}}. \bibinfo{pages}{1--10}.
\newblock


\bibitem[Silva et~al\mbox{.}(2018)]%
        {Silva2018leveraging}
\bibfield{author}{\bibinfo{person}{Nelson Silva}, \bibinfo{person}{Tobias Schreck}, \bibinfo{person}{Eduardo Veas}, \bibinfo{person}{Vedran Sabol}, \bibinfo{person}{Eva Eggeling}, {and} \bibinfo{person}{Dieter~W. Fellner}.} \bibinfo{year}{2018}\natexlab{}.
\newblock \showarticletitle{Leveraging eye-gaze and time-series features to predict user interests and build a recommendation model for visual analysis}. In \bibinfo{booktitle}{\emph{Proceedings of the 2018 ACM Symposium on Eye Tracking Research \& Applications}} (Warsaw, Poland) \emph{(\bibinfo{series}{ETRA '18})}. \bibinfo{publisher}{Association for Computing Machinery}, \bibinfo{address}{New York, NY, USA}, Article \bibinfo{articleno}{13}, \bibinfo{numpages}{9}~pages.
\newblock
\showISBNx{9781450357067}
\href{https://doi.org/10.1145/3204493.3204546}{doi:\nolinkurl{10.1145/3204493.3204546}}


\bibitem[Sinha et~al\mbox{.}(2014)]%
        {sinha2014your}
\bibfield{author}{\bibinfo{person}{Tanmay Sinha}, \bibinfo{person}{Patrick Jermann}, \bibinfo{person}{Nan Li}, {and} \bibinfo{person}{Pierre Dillenbourg}.} \bibinfo{year}{2014}\natexlab{}.
\newblock \showarticletitle{Your click decides your fate: Inferring information processing and attrition behavior from mooc video clickstream interactions}.
\newblock \bibinfo{journal}{\emph{arXiv preprint arXiv:1407.7131}} (\bibinfo{year}{2014}).
\newblock


\bibitem[Smith et~al\mbox{.}(2009)]%
        {smith2009adapting}
\bibfield{author}{\bibinfo{person}{Ray Smith}, \bibinfo{person}{Daria Antonova}, {and} \bibinfo{person}{Dar-Shyang Lee}.} \bibinfo{year}{2009}\natexlab{}.
\newblock \showarticletitle{Adapting the Tesseract open source OCR engine for multilingual OCR}. In \bibinfo{booktitle}{\emph{Proceedings of the International Workshop on Multilingual OCR}}. \bibinfo{pages}{1--8}.
\newblock


\bibitem[Spanjers et~al\mbox{.}(2012)]%
        {Spanjers2012}
\bibfield{author}{\bibinfo{person}{Ingrid A.~E. Spanjers}, \bibinfo{person}{Tamara van Gog}, {and} \bibinfo{person}{Jeroen J.~G. van Merri{\"e}nboer}.} \bibinfo{year}{2012}\natexlab{}.
\newblock \showarticletitle{Segmentation of Worked Examples: Effects on Cognitive Load and Learning}.
\newblock \bibinfo{journal}{\emph{Applied Cognitive Psychology}} \bibinfo{volume}{26}, \bibinfo{number}{3} (\bibinfo{year}{2012}), \bibinfo{pages}{352--358}.
\newblock
\href{https://doi.org/10.1002/acp.1832}{doi:\nolinkurl{10.1002/acp.1832}}


\bibitem[Sweller et~al\mbox{.}(1998)]%
        {sweller1998cognitive}
\bibfield{author}{\bibinfo{person}{John Sweller}, \bibinfo{person}{Jeroen~JG Van~Merrienboer}, {and} \bibinfo{person}{Fred~GWC Paas}.} \bibinfo{year}{1998}\natexlab{}.
\newblock \showarticletitle{Cognitive architecture and instructional design}.
\newblock \bibinfo{journal}{\emph{Educational Psychology Review}}  \bibinfo{volume}{10} (\bibinfo{year}{1998}), \bibinfo{pages}{251--296}.
\newblock


\bibitem[Sweller et~al\mbox{.}(2019)]%
        {Sweller2019}
\bibfield{author}{\bibinfo{person}{John Sweller}, \bibinfo{person}{Jeroen J.~G. van Merri{\"e}nboer}, {and} \bibinfo{person}{Fred Paas}.} \bibinfo{year}{2019}\natexlab{}.
\newblock \showarticletitle{Cognitive Architecture and Instructional Design: 20 Years Later}.
\newblock \bibinfo{journal}{\emph{Educational Psychology Review}} \bibinfo{volume}{31}, \bibinfo{number}{2} (\bibinfo{year}{2019}), \bibinfo{pages}{261--292}.
\newblock
\href{https://doi.org/10.1007/s10648-019-09465-5}{doi:\nolinkurl{10.1007/s10648-019-09465-5}}


\bibitem[Tiellet et~al\mbox{.}(2010)]%
        {tiellet2010design}
\bibfield{author}{\bibinfo{person}{Claudio~AB Tiellet}, \bibinfo{person}{Andr{\'e}~Grahl Pereira}, \bibinfo{person}{Eliseo~Berni Reategui}, \bibinfo{person}{Jos{\'e}~Valdeni Lima}, {and} \bibinfo{person}{Teresa Chambel}.} \bibinfo{year}{2010}\natexlab{}.
\newblock \showarticletitle{Design and evaluation of a hypervideo environment to support veterinary surgery learning}. In \bibinfo{booktitle}{\emph{Proceedings of the 21st ACM Conference on Hypertext and Hypermedia}}. \bibinfo{pages}{213--222}.
\newblock


\bibitem[Torre et~al\mbox{.}(2022)]%
        {torre2022video}
\bibfield{author}{\bibinfo{person}{Ilaria Torre}, \bibinfo{person}{Ilenia Galluccio}, {and} \bibinfo{person}{Mauro Coccoli}.} \bibinfo{year}{2022}\natexlab{}.
\newblock \showarticletitle{Video augmentation to support video-based learning}. In \bibinfo{booktitle}{\emph{Proceedings of the 2022 International Conference on Advanced Visual Interfaces}}. \bibinfo{pages}{1--5}.
\newblock


\bibitem[Vizrt(2022)]%
        {VizLibero}
\bibfield{author}{\bibinfo{person}{Vizrt}.} \bibinfo{year}{2022}\natexlab{}.
\newblock \bibinfo{title}{Viz Libero}.
\newblock \bibinfo{howpublished}{\url{https://www.vizrt.com/products/viz-libero}}.
\newblock
\urldef\tempurl%
\url{https://www.vizrt.com/products/viz-libero}
\showURL{%
\tempurl}


\bibitem[Wang and McCallum(2006)]%
        {wang2006topics}
\bibfield{author}{\bibinfo{person}{Xuerui Wang} {and} \bibinfo{person}{Andrew McCallum}.} \bibinfo{year}{2006}\natexlab{}.
\newblock \showarticletitle{Topics over time: a non-markov continuous-time model of topical trends}. In \bibinfo{booktitle}{\emph{Proceedings of the 12th ACM SIGKDD International Conference on Knowledge Discovery and Data Mining}}. \bibinfo{pages}{424--433}.
\newblock


\bibitem[Wei et~al\mbox{.}(2022)]%
        {wei2022chain}
\bibfield{author}{\bibinfo{person}{Jason Wei}, \bibinfo{person}{Xuezhi Wang}, \bibinfo{person}{Dale Schuurmans}, \bibinfo{person}{Maarten Bosma}, \bibinfo{person}{Fei Xia}, \bibinfo{person}{Ed Chi}, \bibinfo{person}{Quoc~V Le}, \bibinfo{person}{Denny Zhou}, {et~al\mbox{.}}} \bibinfo{year}{2022}\natexlab{}.
\newblock \showarticletitle{Chain-of-thought prompting elicits reasoning in large language models}.
\newblock \bibinfo{journal}{\emph{Advances in Neural Information Processing Systems}}  \bibinfo{volume}{35} (\bibinfo{year}{2022}), \bibinfo{pages}{24824--24837}.
\newblock


\bibitem[Xie et~al\mbox{.}(2017)]%
        {Xie2017}
\bibfield{author}{\bibinfo{person}{Hua Xie}, \bibinfo{person}{Feng Wang}, \bibinfo{person}{Yong Hao}, \bibinfo{person}{Jun Chen}, {et~al\mbox{.}}} \bibinfo{year}{2017}\natexlab{}.
\newblock \showarticletitle{The More Total Cognitive Load is Reduced by Cues, the Better Retention and Transfer of Multimedia Learning: A Meta-Analysis}.
\newblock \bibinfo{journal}{\emph{PLOS ONE}} \bibinfo{volume}{12}, \bibinfo{number}{8} (\bibinfo{year}{2017}), \bibinfo{pages}{e0183884}.
\newblock
\href{https://doi.org/10.1371/journal.pone.0183884}{doi:\nolinkurl{10.1371/journal.pone.0183884}}


\bibitem[Yadav et~al\mbox{.}(2016)]%
        {yadav2016vizig}
\bibfield{author}{\bibinfo{person}{Kuldeep Yadav}, \bibinfo{person}{Ankit Gandhi}, \bibinfo{person}{Arijit Biswas}, \bibinfo{person}{Kundan Shrivastava}, \bibinfo{person}{Saurabh Srivastava}, {and} \bibinfo{person}{Om Deshmukh}.} \bibinfo{year}{2016}\natexlab{}.
\newblock \showarticletitle{Vizig: Anchor points based non-linear navigation and summarization in educational videos}. In \bibinfo{booktitle}{\emph{Proceedings of the 21st International Conference on Intelligent User Interfaces}}. \bibinfo{pages}{407--418}.
\newblock


\bibitem[Zawacki-Richter et~al\mbox{.}(2018)]%
        {zawacki2018research}
\bibfield{author}{\bibinfo{person}{Olaf Zawacki-Richter}, \bibinfo{person}{Aras Bozkurt}, \bibinfo{person}{Uthman Alturki}, {and} \bibinfo{person}{Ahmed Aldraiweesh}.} \bibinfo{year}{2018}\natexlab{}.
\newblock \showarticletitle{What research says about MOOCs--An explorative content analysis}.
\newblock \bibinfo{journal}{\emph{The International Review of Research in Open and Distributed Learning}} \bibinfo{volume}{19}, \bibinfo{number}{1} (\bibinfo{year}{2018}).
\newblock


\bibitem[Zhang et~al\mbox{.}(2022)]%
        {zhang2022towards}
\bibfield{author}{\bibinfo{person}{Gefei Zhang}, \bibinfo{person}{Zihao Zhu}, \bibinfo{person}{Sujia Zhu}, \bibinfo{person}{Ronghua Liang}, {and} \bibinfo{person}{Guodao Sun}.} \bibinfo{year}{2022}\natexlab{}.
\newblock \showarticletitle{Towards a better understanding of the role of visualization in online learning: A review}.
\newblock \bibinfo{journal}{\emph{Visual Informatics}} \bibinfo{volume}{6}, \bibinfo{number}{4} (\bibinfo{year}{2022}), \bibinfo{pages}{22--33}.
\newblock


\bibitem[Zhang et~al\mbox{.}(2023)]%
        {zhang2023visual}
\bibfield{author}{\bibinfo{person}{Huijie Zhang}, \bibinfo{person}{Jialu Dong}, \bibinfo{person}{Cheng Lv}, \bibinfo{person}{Yiming Lin}, {and} \bibinfo{person}{Jinghan Bai}.} \bibinfo{year}{2023}\natexlab{}.
\newblock \showarticletitle{Visual analytics of potential dropout behavior patterns in online learning based on counterfactual explanation}.
\newblock \bibinfo{journal}{\emph{Journal of Visualization}} \bibinfo{volume}{26}, \bibinfo{number}{3} (\bibinfo{year}{2023}), \bibinfo{pages}{723--741}.
\newblock


\bibitem[Zhang et~al\mbox{.}(2019)]%
        {zhang2019scaffomapping}
\bibfield{author}{\bibinfo{person}{Shan Zhang}, \bibinfo{person}{Xiaojun Meng}, \bibinfo{person}{Can Liu}, \bibinfo{person}{Shengdong Zhao}, \bibinfo{person}{Vibhor Sehgal}, {and} \bibinfo{person}{Morten Fjeld}.} \bibinfo{year}{2019}\natexlab{}.
\newblock \showarticletitle{ScaffoMapping: Assisting concept mapping for video learners}. In \bibinfo{booktitle}{\emph{Human-Computer Interaction--INTERACT 2019: 17th IFIP TC 13 International Conference, Paphos, Cyprus, September 2--6, 2019, Proceedings, Part II 17}}. Springer, \bibinfo{pages}{314--328}.
\newblock


\bibitem[Zhao et~al\mbox{.}(2017)]%
        {zhao2017novel}
\bibfield{author}{\bibinfo{person}{Baoquan Zhao}, \bibinfo{person}{Shujin Lin}, \bibinfo{person}{Xiaonan Luo}, \bibinfo{person}{Songhua Xu}, {and} \bibinfo{person}{Ruomei Wang}.} \bibinfo{year}{2017}\natexlab{}.
\newblock \showarticletitle{A novel system for visual navigation of educational videos using multimodal cues}. In \bibinfo{booktitle}{\emph{Proceedings of the 25th ACM International Conference on Multimedia}}. \bibinfo{pages}{1680--1688}.
\newblock


\bibitem[Zhao et~al\mbox{.}(2016)]%
        {zhao2016new}
\bibfield{author}{\bibinfo{person}{Baoquan Zhao}, \bibinfo{person}{Songhua Xu}, \bibinfo{person}{Shujin Lin}, \bibinfo{person}{Xiaonan Luo}, {and} \bibinfo{person}{Lian Duan}.} \bibinfo{year}{2016}\natexlab{}.
\newblock \showarticletitle{A new visual navigation system for exploring biomedical Open Educational Resource (OER) videos}.
\newblock \bibinfo{journal}{\emph{Journal of the American Medical Informatics Association}} \bibinfo{volume}{23}, \bibinfo{number}{e1} (\bibinfo{year}{2016}), \bibinfo{pages}{e34--e41}.
\newblock


\bibitem[Zhou et~al\mbox{.}(2024)]%
        {zhou2024conceptthread}
\bibfield{author}{\bibinfo{person}{Zhiguang Zhou}, \bibinfo{person}{Li Ye}, \bibinfo{person}{Lihong Cai}, \bibinfo{person}{Lei Wang}, \bibinfo{person}{Yigang Wang}, \bibinfo{person}{Yongheng Wang}, \bibinfo{person}{Wei Chen}, {and} \bibinfo{person}{Yong Wang}.} \bibinfo{year}{2024}\natexlab{}.
\newblock \showarticletitle{ConceptThread: Visualizing Threaded Concepts in MOOC Videos}.
\newblock \bibinfo{journal}{\emph{IEEE Transactions on Visualization and Computer Graphics}} (\bibinfo{year}{2024}).
\newblock


\bibitem[Zhu-Tian et~al\mbox{.}(2023)]%
        {Chen2023Sporthesia}
\bibfield{author}{\bibinfo{person}{Chen Zhu-Tian}, \bibinfo{person}{Qisen Yang}, \bibinfo{person}{Xiao Xie}, \bibinfo{person}{Johanna Beyer}, \bibinfo{person}{Haijun Xia}, \bibinfo{person}{Yingcai Wu}, {and} \bibinfo{person}{Hanspeter Pfister}.} \bibinfo{year}{2023}\natexlab{}.
\newblock \showarticletitle{Sporthesia: Augmenting Sports Videos Using Natural Language}.
\newblock \bibinfo{journal}{\emph{IEEE Transactions on Visualization and Computer Graphics}} \bibinfo{volume}{29}, \bibinfo{number}{1} (\bibinfo{year}{2023}), \bibinfo{pages}{918--928}.
\newblock
\href{https://doi.org/10.1109/TVCG.2022.3209497}{doi:\nolinkurl{10.1109/TVCG.2022.3209497}}


\end{thebibliography}








\end{document}